\documentclass[aps,pre,showpacs,showkeys,amsmath,amsfonts,amssymb,superscriptaddress]{revtex4} 
\usepackage[dvips]{graphicx}
\usepackage{bbm}
\usepackage{psfrag}
\newcommand{\beq}{\begin{equation}} 
\newcommand{\eeq}{\end{equation}}
\newcommand{\bea}{\begin{eqnarray}} 
\newcommand{\eea}{\end{eqnarray}}

\input epsf 
 
\begin{document} 
 
\title{Caloric curve of star clusters} 

\author{Lapo Casetti} 
\email{lapo.casetti@unifi.it} 
\affiliation{Dipartimento di Fisica e Astronomia and Centro per lo Studio
delle Dinamiche Complesse (CSDC), Universit\`a di 
Firenze, via G.~Sansone 1, I-50019 Sesto Fiorentino (FI), Italy}  
\affiliation{Istituto Nazionale di Fisica Nucleare (INFN), Sezione di
Firenze, via G.~Sansone 1, I-50019 Sesto Fiorentino (FI), Italy}  
\author{Cesare Nardini} 
\email{cesare.nardini@gmail.com} 
\affiliation{Dipartimento di Fisica e Astronomia and Centro per lo Studio
delle Dinamiche Complesse (CSDC), Universit\`a di 
Firenze, via G.~Sansone 1, I-50019 Sesto Fiorentino (FI), Italy}  
\affiliation{Istituto Nazionale di Fisica Nucleare (INFN), Sezione di
Firenze, via G.~Sansone 1, I-50019 Sesto Fiorentino (FI), Italy}  
\affiliation{Laboratoire de Physique, \'Ecole Normale Sup\'erieure de Lyon, 46, all\'ee d'Italie, F-69007 Lyon, France}  
\date{\today} 
 
\begin{abstract} Self-gravitating systems, like globular clusters or elliptical galaxies, are the prototypes of many-body systems with long-range interactions, and should be the natural arena where to test theoretical predictions on the statistical behaviour of long-range-interacting systems. Systems of classical self-gravitating particles can be studied with the standard tools of equilibrium statistical mechanics, provided the potential is regularized at small length scales and the system is confined in a box. The confinement condition looks rather unphysical in general, so that it is natural to ask whether what we learn with these studies is relevant to real self-gravitating systems. In order to provide a first answer to this question we consider a basic, simple, yet effective model of globular clusters, the King model. This model describes a self-consistently confined system, without the need of any external box, but the stationary state is a non-thermal one. In particular, we consider the King model with a short-distance cutoff on the interactions and we discuss how such a cutoff affects the caloric curve, i.e. the relation between temperature and energy. We find that the cutoff stabilizes a low-energy phase which is absent in the King model without cutoff; the caloric curve of the model with cutoff turns out to be very similar to that of previously studied confined and regularized models, but for the absence of a high-energy gas-like phase. We briefly discuss the possible phenomenological as well as theoretical implications of these results. 
\end{abstract} 
 
\pacs{05.20.-y, 05.70.-a, 05.90.+m, 98.10.+z} 
 
\keywords{Self-gravitating systems; caloric curve; globular clusters} 
 
\maketitle 

\section{Introduction}
\label{sec_intro}

Many objects in the universe, such as globular clusters, elliptical galaxies, or even clusters of galaxies, can be modeled as systems of classical particles mutually interacting via gravitational forces, as long as other interactions are negligible compared to the gravitational ones \cite{BinneyTremaine:book}. From a theoretical point of view, these systems also appear as the prototype of systems with long-range interactions, whose statistical properties have received a considerable attention in the last years \cite{CampaEtAl:physrep,BouchetGuptaMukamel:physicaA2010}. Indeed, the unscreened nature of the gravitational potential makes it truly long-range in any situation and should make self-gravitating systems the ideal candidates where to test theoretical predictions on the equilibrium and non-equilibrium statistical behaviour of many-particle systems with long-range interactions, like the presence of negative microcanonical specific heats, non-homogeneous equilibrium states, long-lived quasi-stationary states \cite{CampaEtAl:physrep}.

However, as far as equilibrium properties are concerned, self-gravitating systems suffer from a serious drawback: they do not have ``true'' thermal equilibrium states, neither in the canonical nor in the microcanonical ensemble. This lack of thermal equilibrium is due to two main reasons (see e.g.\ \cite{IspolatovCohen:prl2001,IspolatovCohen:pre2001,BinneyTremaine:book,Choudhouri:book,Bertin:book} for more detailed discussions). $(i)$ The gravitational potential is singular for vanishing distance between two particles, so that entropy and free energy are infinite for a system of more than two particles \cite{Padmanabhan:physrep}: physically this means that, at equilibrium, (at least part of) the system would collapse in states with infinite density. $(ii)$ If the particles have a thermal (i.e., Maxwellian) velocity distribution they will tend to escape the system since the escape velocity is finite unless the mass of the system itself is infinite. Equivalently, a thermal velocity distribution in a self-gravitating system in unbounded space is possible only if the system has infinite mass \cite{Choudhouri:book}.

From a physical point of view the first problem admits a simple solution. No real system exists where the only non-negligible interaction is classical gravity at {\em all} length scales: either the interacting ``particles'' are macroscopic bodies like planets or stars or galaxies, or quantum effects must be taken into account below a certain length scale. In both cases, a length scale exists (the size of the bodies or the scale where quantum effects set in) below which the potential has no longer the classical gravitational form. If one is not interested in small-scale details, the potential can be regularized e.g.\ by replacing it with a softened one at short distances (see e.g.\ \cite{Padmanabhan:physrep,ChavanisIspolatov:pre2002}) or by directly considering self-gravitating fermions \cite{ChavanisIspolatov:pre2002}.

The second problem is tougher, and one is (somehow artificially) forced to confine the system in a box so that statistical equilibrium exists. On physical grounds such an approximation may be reasonable in those cases where the evaporation rate is slow compared to the other time scales in the system \cite{FanelliMerafinaRuffo:pre2001}, and this happens in many astrophysical systems \cite{BinneyTremaine:book,Maoz:ApJL1998}; however, in some situations it may also induce nonphysical features, as we shall see in the following. We note that also when computing the entropy or free energy of a fluid or a gas one encloses it in a box. However, not only in that case this is precisely what is done in a laboratory---at variance with the case of a galaxy or of a star cluster---but, being the interactions short-ranged and the equilibrium states homogeneous, the size, the shape and the position of the container have no effect on the bulk properties of the fluid. This is not the case for self-gravitating systems, the forces being long-ranged and the equilibrium states non-homogeneous.

Once regularized and confined, self-gravitating systems do have thermal equilibria, both in the canonical and in the microcanonical ensemble \cite{Kiessling:jsp1989,Kiessling:rmp2009}, so that they can be approached with the standard tools of equilibrium statistical mechanics, like mean-field theory, study of simplified models, and numerical simulations. 
In the last decades a lot of models, ranging from minimalistic toy models to more realistic ones, have been studied from this point of view (we shall discuss some of these results, referring to the original papers, in Sec.\ \ref{Sec_sg_eq}: for a review see e.g.\ \cite{Padmanabhan:physrep,Chavanis:ijmpb2006}).
These studies have remarkably shown that the collective behaviour of all these models in the microcanonical ensemble is qualitatively almost the same. At high energy there is a gas-like phase; lowering the energy a negative specific heat phase, where temperature grows while decreasing the energy, is found; this is the phase dominated by gravity. At even lower energies the specific heat becomes positive again and remains positive until the minimum energy is eventually reached. In this low-energy phase the physics is dominated by the regularization. Also the canonical behaviour of these systems is essentially the same and is much simpler than the microcanonical one: there is a discontinuous phase transition from the high-temperature gas-like phase to the collapsed phase dominated by regularization. An interval in energy, containing the region where the specific heat is negative, is avoided in the canonical ensemble. This is a clear manifestation of ensemble inequivalence.

It is important to stress that even if equilibrium states exist, nothing guarantees that they are actually reached by the dynamical evolution of self-gravitating systems regardless of the initial conditions: indeed, long-range interacting systems have long-lived off-equilibrium states (called quasi-stationary states, QSSs) where the dynamics can remain trapped for times rapidly growing with the number of degrees of freedom \cite{CampaEtAl:physrep} so that relaxation to thermal equilibrium (if any) may take extremely long times (for a study of this phenomenon in a toy model of a gravitating system see \cite{JoyceWorrakitpoonpon:pre2011}). This, in addition to the non-physical assumption of confinement, might suggest the study of equilibrium properties of self-gravitating systems has only a purely academic interest as a mathematical problem. However, real self-gravitating systems such as globular clusters or elliptical galaxies do show remarkably similar collective properties \cite{Bertin:book} suggesting that some kind of relaxation has occurred. This makes at least reasonable, although not firmly and rigorously grounded, a study of self-gravitating systems from an equilibrium point of view. 

Then, a question naturally arises: Is what we learned with the above mentioned studies of regularized and confined models relevant to real self-gravitating systems, where no confining box is present? The aim of the present paper is to provide a first, preliminary answer to this question by studying the caloric curve of a basic, simple, yet effective model of globular clusters, the King model \cite{King:aj1966_III}. This model describes a self-consistently confined system, without the need of any external box, but the stationary state is a non-thermal one. In particular, we shall consider the King model with a short-distance cutoff on the interactions and we shall discuss how such a cutoff affects the caloric curve. 

The paper is organized as follows. In Sec.\ \ref{Sec_sg_eq} we briefly review the known results on the equilbrium statistical mechanics of self-gravitating systems relevant to the understanding of our work. Section \ref{Sec_King} contains a discussion of the King model of star clusters as well as a calculation of its caloric curve. Section \ref{Sec_King_cutoff} is devoted to the King model with a short-distance cutoff and contains the main results of the present work, especially concerning the caloric curve of that model and its dependence on the cutoff. Finally, Sec.\ \ref{Sec_final} is devoted to some concluding remarks.

\section{Equilibrium statistical mechanics of self-gravitating systems}
\label{Sec_sg_eq}

Let us consider a system of $N$ classical point particles of mass $m$, mutually interacting via gravitational forces. The Hamiltonian, written as a function of the positions $\mathbf{r}_i$ and of the velocities $\mathbf{v}_i$ of the particles, is
\beq
{\cal H}\left({\bf {r}}_1,\ldots,{\bf {r}}_N,{\bf {v}}_1,\ldots,{\bf {v}}_N \right) = \frac{m}{2}\sum_{i=1}^N {v}_i^2 - G m^2 \sum_{i=1}^N \sum_{j > i}^N \frac{1}{\left|{\bf {r}}_i - {\bf {r}}_j\right|} \, ,
\label{H}
\eeq
where $v_i = \left|\mathbf{v}_i \right|$ and $G$ is Newton's gravitational constant. The equilibrium properties of such a system at constant energy $E$ can be derived from the microcanonical (Boltzmann) entropy\footnote{Throughout the paper we set Boltzmann's constant to unity so that temperatures are measured in energy units.}
\beq
S(E) = \log \omega(E)
\label{S}
\eeq
where $\omega(E)$ is the density of states, defined by
\beq
\omega(E) = \int \delta\left[\mathcal{H}\left({\bf {r}}_1,\ldots,{\bf {r}}_N,{\bf {v}}_1,\ldots,{\bf {v}}_N \right) - E \right] d\Gamma\, ,
\label{omega}
\eeq
with $d\Gamma = \prod_{i=1}^N d\mathbf{r}_i\,d\mathbf{v}_i$. Equilibrium properties at fixed temperature $T$ stem from the Helmholtz free energy
\beq
F(T) = - T \log Z(T)
\label{F}
\eeq
where $Z(T)$ is the canonical partition function given by 
\beq
Z(T) = \int \exp\left[-\beta\mathcal{H}\left({\bf {r}}_1,\ldots,{\bf {r}}_N,{\bf {v}}_1,\ldots,{\bf {v}}_N \right)\right] d\Gamma\, ,
\label{Z}
\eeq
where $\beta = T^{-1}$. As already discussed in Sec.\ \ref{sec_intro}, the integrals in Eqs.\ (\ref{omega}) and (\ref{Z}) are well-defined only when the interaction potential is regularized at small length scales and the system is enclosed in a finite volume $V$.

\subsection{The isothermal sphere}
\label{sec_isosphere}
The analytical calculation of $\omega(E)$ or $Z(T)$ for any system (\ref{H}) with $N > 2$ is not feasible. The simplest approach to the equilibrium statistical mechanics of such a system when $N$ is large is the mean-field approximation, which becomes exact in the $N\to\infty$ limit \cite{CampaEtAl:physrep,Padmanabhan:physrep,Chavanis:ijmpb2006}. The mean-field approach can even be pursued without the introduction of a short-distance regularization: in this case, the mean-field physical quantities remain finite, although their derivation from Eqs.\ (\ref{S}) and (\ref{F}) is somehow ill-defined (see e.g.\ \cite{Padmanabhan:physrep}). As we shall see in the following, however, neglecting the regularization does have a big effect on the physics.

Let us now briefly recall an important example of the mean-field approach to the equilibrium of self-gravitating systems, the so-called ``isothermal sphere'', first studied in a seminal work by Antonov \cite{Antonov:1962}. This will also allow us to set up some notation which will be useful in the following. The system is enclosed in a spherical box of radius $R$ and its state is described by a single-particle distribution function $f(\mathbf{r},\mathbf{v})$ such that the total mass $M$ is
\beq
M = \int f(\mathbf{r},\mathbf{v}) \, d\mathbf{r}\,d\mathbf{v}\, ,
\label{M}
\eeq
and its microcanonical entropy is a functional of $f$ given by 
\beq
S[f] = - \int \frac{f(\mathbf{r},\mathbf{v})}{f_0} \log \left[\frac{f(\mathbf{r},\mathbf{v})}{f_0}\right]\, d\mathbf{r}\,d\mathbf{v}\,  ,
\label{mfentropy}
\eeq
where $f_0$ is a constant such that $f/f_0$ is dimensionless, and otherwise arbitrary. Microcanonical equilibrium states correspond to maxima of the entropy. Here we have {\em not} introduced any short-range regularization of the potential, so that strictly speaking the entropy functional (\ref{mfentropy}) does not have a global maximum corresponding to a smooth distribution function $f$, because a singular, collapsed state would have a larger (actually infinite) entropy \cite{IspolatovCohen:pre2001,Chavanis:ijmpb2006}. However, one can look for local maxima of the entropy and consider them as our equilibrium states, although they are only metastable states. Imposing the stationarity condition on the entropy one finds spherically symmetric distributions, i.e., $f(\mathbf{r},\mathbf{v}) = f(r,v)$, which depend only on the single-particle energy:
\beq
f(r,v) = C \exp\left\{-\gamma \left[ \frac{v^2}{2} + \varphi( r ) \right]\right\}
\label{f}
\eeq
where $C$ is a normalization constant related to the total mass $M$, $\gamma$ is proportional to the inverse temperature and $\varphi( r )$ is the mean-field gravitational potential which obeys the Poisson equation
\beq
\nabla^2 \varphi( r ) = 4\pi G \varrho( r ) \, ,
\label{poisson}
\eeq
where $\varrho$ is the mass density 
\beq
\varrho(r )  = \int f(r,v)  \, d\mathbf{v}\, .
\label{mfdensity}
\eeq
Finding the $f$ corresponding to the local extrema of the entropy amounts then to solving the Poisson equation (\ref{poisson}), which becomes effectively one-dimensional due to the spherical symmetry, coupled to Eq.\ (\ref{f}) via Eq.\ (\ref{mfdensity}). This cannot be done analytically but is an easy task on a computer. 

The collective behaviour in the microcanonical ensemble is conveniently encoded in the caloric curve, i.e., the relation between the temperature $T$ and the energy $E$. To define these quantities we observe that the mean-field kinetic energy $K$ is given by 
\beq
K = \frac{1}{2} \int v^2 f(r,v) \, d\mathbf{r}\,d\mathbf{v}\,  
\label{mfkin}
\eeq
and is related to the temperature $T = (dS/dE)^{-1}$ via 
\beq
K = \frac{3T}{2}~,
\label{T}
\eeq
while the mean-field potential energy $U$ is 
\beq
U = 
\frac{1}{2} \int \varrho( r ) \varphi( r ) \, d\mathbf{r}\,   ,
\label{mfpot}
\eeq
so that $E = K + U = \frac{3T}{2} + U$. Introducing dimensionless energy and temperature as
\bea
\varepsilon & = & \frac{RE}{GM^2} \label{eps}\\
\vartheta & = & \frac{RT}{GM^2} \label{theta}
\eea
the caloric curve is obtained by plotting $\vartheta$ as a function of $\varepsilon$.
\begin{figure}
\psfrag{t}{$\vartheta$}
\psfrag{e}{$\varepsilon$}
\psfrag{-0.2}{\hspace{-0.2cm}$-0.2$}
\psfrag{0.0}{$0.0$}
\psfrag{0.2}{$0.2$}
\psfrag{0.4}{$0.4$}
\psfrag{0.5}{$0.5$}
\psfrag{0.6}{$0.6$}
\psfrag{0.7}{$0.7$}
\includegraphics[width=12cm,clip=true]{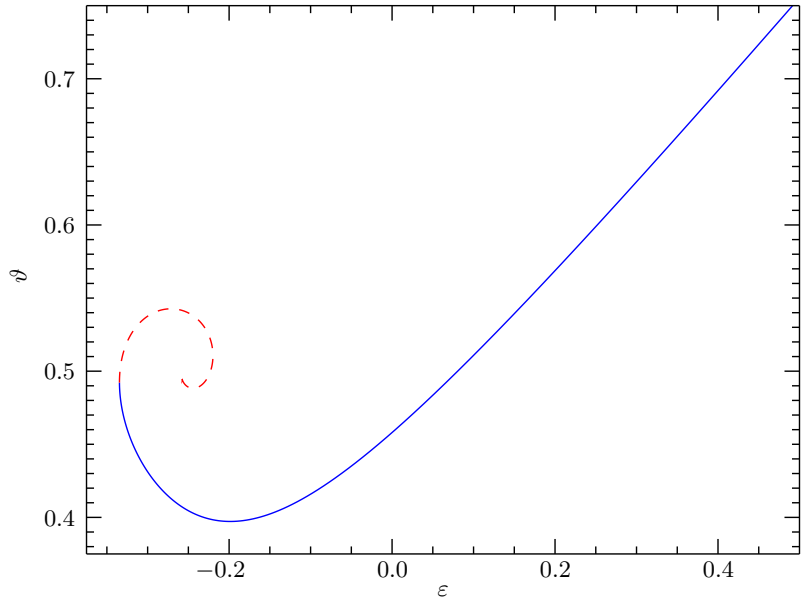}
\caption{Caloric curve of the isothermal sphere. The red dashed line corresponds to unstable states.}
\label{figure:isosphere}
\end{figure}
The caloric curve of the isothermal sphere is reported in Fig.\ \ref{figure:isosphere}. The density contrast $\varrho(0)/\varrho( R )$ grows along the curve, starting from $\varrho(0)/\varrho( R ) = 1$ at $\varepsilon = +\infty$. At high energies $\vartheta$ is approximately proportional to $\varepsilon$ and the systems is in a gas-like phase. By decreasing the energy, the temperature reaches a minimum value $\vartheta^{\text{iso}}_{\text{min}}\simeq 0.397$ and then begins to grow, so that the specific heat becomes negative. A minimum value of the energy $\varepsilon^{\text{iso}}_{\text{min}}\simeq -0.335$, corresponding to a density contrast $\varrho(0)/\varrho( R ) \simeq 709$, is eventually reached. The caloric curve then starts spiralling until a singular sphere is reached, with infinite density contrast. However, all the states with density contrast larger than 709 are unstable, because they are saddle points of the entropy rather than local maxima \cite{Padmanabhan:physrep,Chavanis:ijmpb2006}. The part of the caloric curve corresponding to unstable states is plotted as a dashed line in Fig.\ \ref{figure:isosphere}.  

The presence of a minimal energy $\varepsilon_{\text{min}}$ below which no equilibria are present is customarily interpreted as the ``gravothermal catastrophe'': when $\varepsilon < \varepsilon^{\text{iso}}_{\text{min}}$ the system would collapse to a singular state. However, the presence of the gravothermal catastrophe is a consequence of the fact that we did not regularize the interactions at small length scales: as we shall see in the following, the introduction of a short-distance cutoff on the interaction potential removes the gravothermal catastrophe and stabilizes a low-energy phase. 
 
\subsection{Confined models with a short-distance cutoff}
\label{sec_modelcutoff}
As already discussed in Sec.\ \ref{sec_intro}, a short-distance cutoff on the interactions must be introduced not only to make the density of states (\ref{omega}) and the partition function (\ref{Z}) well-defined\footnote{The small-scale regularization is necessary for the three-dimensional gravitational potential; for lower-dimensional systems, where the gravitational potential is a solution of the Poisson equation in two or in one dimension, such a regularization is not strictly required. Examples of models of gravitating systems in less than three dimensions are the sheet model \cite{HohlFeix:apj1967} and the self-gravitating gas on the $\mathbb{S}^2$ sphere \cite{Kiessling:jstat2011}.}, but also because it is physically motivated: no real system exists where the interaction potential is the classical gravitational one at {\em all} length scales.  

In the last decades, equilibrium statistical mechanics has been studied for many models of self-gravitating systems, defined in a confined volume and with small-scale regularization of the potential. Without entering the details, let us recall some of the latter models in order to underline their many common features, despite the models may be very different from each other. 

The isothermal sphere with cutoff, i.e., the model discussed above with the potential modified by the introduction of a small-scale cutoff, was first considered, in an approximate way, in \cite{AronsonHansen:apj1972}; more recently it has been fully studied numerically, see e.g.\ \cite{IspolatovCohen:prl2001,IspolatovCohen:pre2001} (also for $1/r^\alpha$ potentials with $\alpha \not = 1$) and \cite{Chavanis:ijmpb2006,ChavanisIspolatov:pre2002}. In these works the regularization of the potential is achieved via the so-called Plummer softening: the potential energy is given by 
\beq
{\cal V}\left({\bf {r}}_1,\ldots,{\bf {r}}_N\right) = \sum_{i=1}^N \sum_{j > i}^N v_a\left(\left|{\bf {r}}_i - {\bf {r}}_j\right|\right) 
\label{reg_potential}
\eeq
with
\beq
v_a(x) = - \frac{Gm^2}{\sqrt{x^2 + a^2}} \, ,
\label{cutoff}
\eeq
where $a$ is the small-scale cutoff. The isothermal sphere with cutoff is a mean-field model, but the statistical mechanics of $N$ particles mutually interacting with the potential energy given by Eq.\ (\ref{reg_potential}) has been studied also via microcanonical and canonical Monte Carlo simulations \cite{DeVegaSanchez:npb2002}. A different kind of regularization has also been considered, i.e., a basis function regularization: the softening is achieved by truncating to $n$ terms an expansion of the Newtonian potential in spherical Bessel functions \cite{FollanaLaliena:pre2000} (a similar regularization was used in $N$-particle dynamical simulations \cite{CerrutiCiprianiPettini:mnras2001}). Also self-gravitating fermions have been studied, where the cutoff is not imposed but is a consequence of the Pauli exclusion principle \cite{Chavanis:ijmpb2006,ChavanisIspolatov:pre2002}.

As is customary in statistical mechanics, many simplified models have been proposed which try to grasp the essential physics by simplifying either the interactions or the geometry or both. A really minimalistic model was proposed by Thirring in the 1970's \cite{Thirring:zfp1970}, where the only feature of the gravitational interactions that is retained is nonadditivity. Other popular models where the geometry is very simplified are the shell model \cite{YoungkinsMiller:pre2000} and the self-gravitating ring (SGR) \cite{SotaEtAl:pre2001,TatekawaEtAl:pre2005}; from the latter a minimalistic model of the same kind as Thirring's was recently derived \cite{jstat2010}, showing that 
minimalistic models can be accurate also at a quantitative, and not only 
qualitative, level. Another simple model is the binary star model first introduced by Padmanabhan (see \cite{Padmanabhan:physrep} and \cite{Chavanis:ijmpb2006}), where only two particles are considered.

\begin{figure}
\includegraphics[width=12cm,clip=true]{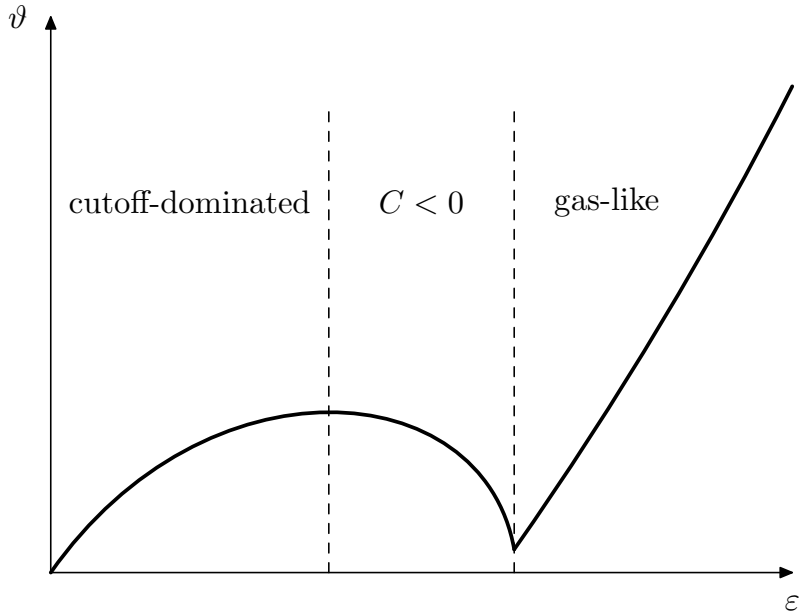}
\caption{Sketch of a typical caloric curve of a model of a regularized and confined self-gravitating system.}
\label{figure:caloric}
\end{figure}

All the above discussed models, although very different from each other, do share two main features: they are confined, in the sense that the configuration space is a compact manifold (either without boundary, as in the case of the SGR model, or with boundary when the systems is defined in $\mathbb{R}^3$ and confined in a box), and the potential is regularized at small length scales. The detailed properties of these models can be very different and also depend on the cutoff scale, and we refer the reader to the original works for these details. This notwithstanding, as anticipated in Sec.\ \ref{sec_intro}, when the cutoff length is sufficiently small the caloric curve of all these models is qualitatively almost the same. At high energy there is a gas-like phase, where the caloric curve is close to a straight line with positive slope. Lowering the energy the specific heat becomes negative. In some cases, the gas-like and the negative specific heat regions are separated by a microcanonical phase transition whose order does depend on the details of the model and on the specific short-distance regularization chosen. In other cases, a transition occurs within the negative specific heat region, i.e., there is a phase transition between two different negative specific heat phases\footnote{In some cases the presence of a true transition with a singularity in the thermodynamic function can not be proved, however available data suggest that at least a very sharp crossover exists.}. In these cases, the phase transition can be interpreted as the counterpart of the gravothermal catastrophe for models with short-range cutoff. The negative specific heat region, where temperature grows while decreasing the energy, is dominated by gravity; here the equilibrium state is clustered but typically a dense cluster is accompanied by a diffuse halo. At even lower energies the caloric curve reaches a maximum and the curve bends down, with a specific heat which becomes positive again and remains positive until the minimum energy is eventually reached. In this low-energy phase the physics is dominated by the regularization and the equilibrium state is strongly clustered. We note that, at variance with the isothermal sphere, the minimum energy of these models is the true minimum of their potential energy. A typical caloric curve of a model of a regularized and confined self-gravitating system is sketched in Fig.\ \ref{figure:caloric}. 

In the canonical ensemble, the intermediate negative specific heat phase is suppressed and there is a discontinuous phase transition between the gas-like and the low-energy collapsed phase.

The general conclusion that can be drawn from these studies is that the introduction of a short-distance cutoff stabilizes a low-energy phase, which is the continuation of the negative specific heat region of the isothermal sphere. In this region the specific heat remains negative until eventually the regularization of the potential becomes dominant and temperature starts to decrease with energy until the minimum is reached: the smaller the cutoff length, the larger the negative specific heat region is. The gravothermal catastrophe in the microcanonical ensemble disappears. 

However, the presence of a high-energy gas-like phase, where the equilibrium state is nearly homogeneous, is a consequence of the confinement of the system. Hence we do not expect such phase to be observable in real self-gravitating system, except for some special cases where the system is indeed confined in a restricted volume by dynamical or other constraints\footnote{Examples of this kind may be self-gravitating objects like the 
rings of Saturn, in which the dynamical constraint is given by a balance 
between the rotation and the attraction exerted by the planet.}. Clearly, also a sharp transition between the gas-like phase and the negative specific heat phase would not be observable, while a transition between different collapsed phases would not be ruled out. More generally it is unclear whether the ``universal'' collective behaviour observed in models of confined and regularized systems does indeed resemble the behaviour of real self-gravitating systems. In the next sections, and especially in Sec.\ \ref{Sec_King_cutoff}, we shall address this problem by studying the caloric curve of a model with small distance cutoff but no confining box. Such a model is a modification of a classic model of globular clusters, the dynamical King model \cite{King:aj1966_III}: hence the next Section is devoted to a brief introduction to this model from the point of view of statistical mechanics.

\section{King model of star clusters}
\label{Sec_King}
Globular clusters are systems of $10^5 \div 10^6$ stars, orbiting galaxies. Nearly 150 globular clusters orbiting the Milky Way have been discovered to date; around 500 globular clusters are estimated for the Andromeda galaxy M31 and roughly $10^4$ for the giant elliptical galaxy M87 in the Virgo cluster \cite{BinneyMerrifield:book}. Their shape is spherical to a very good degree of approximation and their size is finite: the limiting radius is referred to as ``tidal radius'' (denoted by $r_t$) because for $r > r_t$ the tidal effects due to the host galaxy gravitational field dominate over the clusters's own gravity. Globular clusters are believed to be composed almost exclusively by stars, without relevant amounts of dust or gas (and dark matter too). This makes globular clusters almost ideal self-gravitating systems. Moreover, they are very old: the age of globular clusters is typically larger than $10^{10}$ years \cite{BinneyMerrifield:book}, which is also the order of magnitude of the collisional relaxation time as 
estimated via the classical Chandrasekhar formula, so that it is reasonable that they have undergone some kind of relaxation.

Globular clusters are therefore the natural choice as examples of real self-gravitating systems that could be approached by statistical mechanics. However, standard thermal equilibrium can not be present in these systems, since they are not enclosed in a box: their velocity distribution will not be a pure Maxwellian one. However, it is reasonable to assume the velocity distribution is as close to a Maxwellian as possible. Indeed, the idea at the basis of the King model is a modification of the isothermal sphere discussed in Sec.\ \ref{sec_isosphere} (see e.g.\ \cite{BinneyTremaine:book}). The state of the system is defined, as in the isothermal sphere, by a spherically symmetric single-particle distribution function $f({r},{v})$ such that the total mass $M$ is given by Eq.\ (\ref{M}). The velocity distribution is obtained by ``lowering a Maxwellian'', such that no particle can have a velocity larger than the escape velocity $v_e ( r )$. The King distribution function is \cite{King:aj1966_III}
\beq
f(r,v) = \left\{
\begin{array}{ll}
C\, e^{-2\gamma\varphi( r ) } \left[e^{-\gamma v^2} - e^{-\gamma v_e^2 ( r ) }\right] & \text{if}~ v^2 < v_e^2 ( r )\, ,\\
0 & \text{if}~ v^2 \geq v_e^2 ( r ) \, ,
\end{array}
\right.
\label{King_f}
\eeq
where $C$ is proportional to the total mass $M$ and $\gamma$ is an inverse potential scale which is no longer directly proportional to the inverse temperature as the $\gamma$ of the isothermal sphere was, since the velocity distribution function is no longer Gaussian. Let us choose energies such that $\varphi(r_t) = 0$, with the tidal radius $r_t$ defined as the radius such that the density $\varrho$, still given by Eq.\ (\ref{mfdensity}), vanishes; then $v^2_e ( r ) = - 2\varphi( r )$. 
Given these definitions, everything goes on as in the case of the isothermal sphere; since we have not introduced any short-distance cutoff on the iteractions (we shall consider this case in Sec.\ \ref{Sec_King_cutoff}) the potential $\varphi( r )$ obeys the Poisson equation (\ref{poisson}) and we get an effectively one-dimensional differential equation whose solution gives the explicit $f$. The difference with the case of the isothermal sphere is that the $f$ thus obtained is no longer a stationary point of the microcanonical entropy\footnote{Clearly one could define an ``effective entropy'' by inserting the King distribution function (\protect\ref{King_f}) into Eq.\ (\protect\ref{mfentropy}), but the latter could hardly be given the meaning of a thermodynamic entropy. See also the discussion in Sec.\ \protect\ref{Sec_final}.}. In practice one chooses a value for the central potential energy $\varphi_0$ or for the central density $\varrho_0 = \varrho(0)$ and integrates the differential equation to get the density profile $\varrho( r )$; we refer the reader to the original paper \cite{King:aj1966_III} and to \cite{BinneyTremaine:book,McLaughlinEtAl:apjs2006} for the details. If $r_t \to \infty$, then the King model becomes identical to an isothermal sphere with $R \to \infty$ and $M \to \infty$. 

Assuming that ``light follows mass'', i.e., that the luminosity density is proportional to the mass density, one can project $\varrho ( r ) $ onto the plane of the sky \cite{BinneyTremaine:book,King:aj1966_III} to compare the result with photometric observations. The resulting distributions provide very good fits\footnote{King actually proposed an empirical density law to fit the observed profiles in \cite{King:aj1962_I}, and only some years later found that a very similar density profile could result from the physically motivated model proposed in \cite{King:aj1966_III}.} to the observed densities of nearly 80\% of the Milky way globular clusters \cite{King:aj1966_III,PetersonKing:aj1975_VI,TragerKingDjorgovski:aj1995}. Also the internal kinematics (the velocity dispersion profile) of the globular clusters is reasonably well described by the King model \cite{PetersonKing:aj1975_VI,McLaughlinVanDerMarel:apjs2005}, even if there may be significant deviations (see e.g.\ \cite{ZocchiBertinVarri:preprint2012}). Although the recent availability of higher-quality data has shown that the King model does not provide an optimal description of globular clusters, it is still considered the best physically motivated model to date (see e.g.\ \cite{McLaughlinVanDerMarel:apjs2005}) and it remains the standard data interpretation and classification tool in this field \cite{McLaughlinEtAl:apjs2006}.

\subsection{Caloric curve}
\label{Sec_caloric_King}

The kinetic and potential energy of the King model can be calculated as in the case of the isothermal sphere. Kinetic energy is still given by  Eq.\ (\ref{mfkin}), while potential energy is given by the same expression as in Eq.\ (\ref{mfpot}), provided the potential is not the potential $\varphi( r )$ entering Eq.\ (\ref{King_f}) but the potential vanishing at infinity, $\varphi_\infty( r ) = \varphi(r ) - \varphi(\infty) = \varphi(r ) - GM/r_t$; the potential energy of the King model can thus be written as
\beq
U = 
-\frac{GM^2}{2r_t} + 2\pi \int_0^{r_t} \varrho( r )\, \varphi( r ) \, r^2 d{r}\, .
\label{King_pot}
\eeq
Then, the temperature $T$ may still be defined as in Eq. (\ref{T}), although we do no longer have a microcanonical entropy such that $T = (dS/dE)^{-1}$. Hence a ``microcanonical caloric curve'' is still given by plotting temperature versus energy, although strictly speaking there is no equilibrium microcanonical ensemble; the constraint of constant energy, however, justifies the analogy. It remains to define the dimensionless energy $\varepsilon$ and temperature $\vartheta$, and in close analogy to the case of the isothermal sphere we set 
\bea
\varepsilon & = & \frac{r_t E}{GM^2} \label{King_eps}\\
\vartheta & = & \frac{r_t T}{GM^2} \label{King_theta}
\eea
where the tidal radius $r_t$ plays the r\^ole of the radius $R$ of the box of the isothermal sphere. Using $r_t$ as unit of lengths is not customary in the context of the King model, where a length scale derived from dimensional analysis of the Poisson equation is typically used \cite{King:aj1966_III,McLaughlinVanDerMarel:apjs2005}; however, our choice simplifies the comparison between the King model and the isothermal sphere, because one is somehow using the same unit of length in the two cases.

At variance with the isothermal sphere, the shape of the caloric curve $\vartheta(\varepsilon)$ of the King model can be predicted without performing any calculation: since the King model gives a stationary distribution function and involves only pure gravitational interactions, without external containers or short-distance cutoffs, the virial theorem explicitly gives the relation between temperature and energy as
\beq
\vartheta = - \frac{2}{3} \varepsilon~.
\label{King_caloric_curve}
\eeq
However, the virial theorem alone does not tell anything on the boundaries (if any) of the domain of the caloric curve. From Eq.\ (\ref{King_pot}) we have that the dimensionless potential energy of the King model
\beq
u = \frac{r_t U}{GM^2}
\eeq
is surely bounded above by $u = -1/2$, but the calculations show that the real upper bound is smaller, and turns out to be $u^{\text{King}}_{\text{max}} \simeq -1.20$.  Hence, using again the virial relation $K = -U/2$ we have that $\varepsilon \leq \varepsilon^{\text{King}}_{\text{max}} \simeq -0.60$ in the King model. Performing the calculations shows that also a lower bound on energy exists, i.e., $\varepsilon \in [\varepsilon^{\text{King}}_{\text{min}}, \varepsilon^{\text{King}}_{\text{max}}]$ where 
\beq
\left\{ \begin{array}{ccc}
\varepsilon^{\text{King}}_{\text{min}} & \simeq & -2.13\, , \\
 & & \\
\varepsilon^{\text{King}}_{\text{max}} & \simeq & -0.60\, . 
\end{array} \right.
\label{King_emax}
\eeq
The latter result means that no King distribution (\ref{King_f}) exists such that the dimensionless mean-field energy is larger than $\varepsilon^{\text{King}}_{\text{max}}$ or smaller than $\varepsilon^{\text{King}}_{\text{min}}$.
Due to the virial relation (\ref{King_caloric_curve}) also the temperature is bounded, i.e., $\vartheta \in [\vartheta^{\text{King}}_{\text{min}} ,\vartheta^{\text{King}}_{\text{max}}]$ with 
\beq
\left\{ \begin{array}{ccccc}
\vartheta^{\text{King}}_{\text{min}} & = & -\frac{2}{3} \varepsilon^{\text{King}}_{\text{max}} & \simeq & 0.40\, ,\\
 & & & & \\
\vartheta^{\text{King}}_{\text{max}} & = & -\frac{2}{3} \varepsilon^{\text{King}}_{\text{min}} & \simeq & 1.42\, .
\end{array} \right.
\label{King_tmax}
\eeq
This is at variance with the isothermal sphere, where the temperature is unbounded above due to the existence of the gas phase; but if we consider only the negative specific heat region of the isothermal sphere, also in that case the dimensionless temperature is bounded from above an below; incidentally, the two lower bounds are very similar. Hence, the qualitative appearance of the caloric curve of the King model is not very different from that of the isothermal sphere, provided one removes in the latter the gas-like phase due to the presence of the container and one takes into account the virial theorem constraint; the King model minimal energy $\varepsilon^{\text{King}}_{\text{min}} \simeq -2.13$, is, however, sizably smaller than the corresponding value of the isothermal sphere $\varepsilon^{\text{iso}}_{\text{min}}\simeq -0.335$.

\begin{figure}
\psfrag{t}{$\vartheta$}
\psfrag{e}{$\varepsilon$}
\psfrag{-2.5}{\hspace{-0.2cm}$-2.5$}
\psfrag{-2.0}{\hspace{-0.2cm}$-2.0$}
\psfrag{-1.5}{\hspace{-0.2cm}$-1.5$}
\psfrag{-1.0}{\hspace{-0.2cm}$-1.0$}
\psfrag{-0.5}{\hspace{-0.2cm}$-0.5$}
\psfrag{0.2}{$0.2$}
\psfrag{0.4}{$0.4$}
\psfrag{0.6}{$0.6$}
\psfrag{0.8}{$0.8$}
\psfrag{1.0}{$1.0$}
\psfrag{1.2}{$1.2$}
\psfrag{1.4}{$1.4$}
\psfrag{1.6}{$1.6$}
\includegraphics[width=12cm,clip=true]{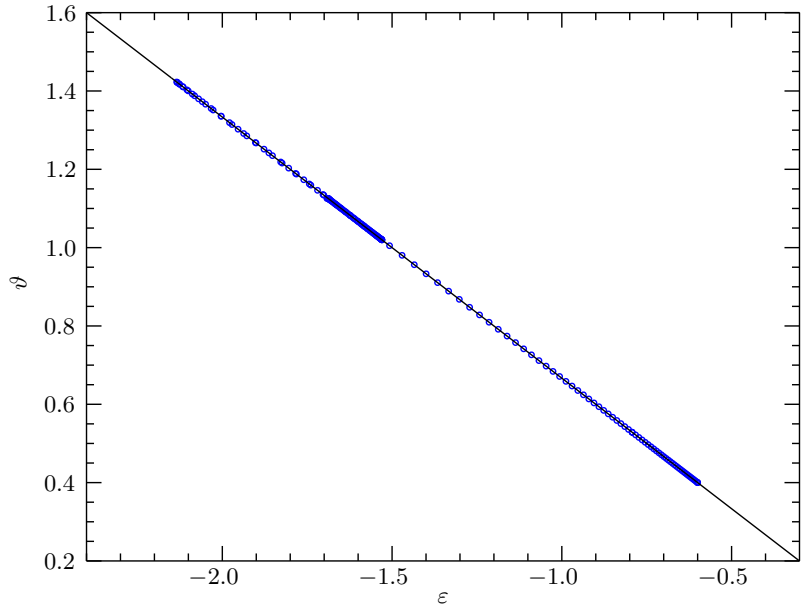}
\caption{Caloric curve of the King model. The solid line is the $\vartheta = -\frac{2}{3} \varepsilon$ virial law and blue circles are the values of $\vartheta$ and $\varepsilon$ calculated using King distribution functions.}
\label{figure:King_caloric}
\end{figure}

\begin{figure}
\psfrag{t0}{$\vartheta_0$}
\psfrag{e}{$\varepsilon$}
\psfrag{-2.0}{\hspace{-0.2cm}$-2.0$}
\psfrag{-1.5}{\hspace{-0.2cm}$-1.5$}
\psfrag{-1.0}{\hspace{-0.2cm}$-1.0$}
\psfrag{-0.5}{\hspace{-0.2cm}$-0.5$}
\psfrag{0.5}{$0.5$}
\psfrag{1.0}{$1.0$}
\psfrag{1.5}{$1.5$}
\psfrag{2.0}{$2.0$}
\includegraphics[width=12cm,clip=true]{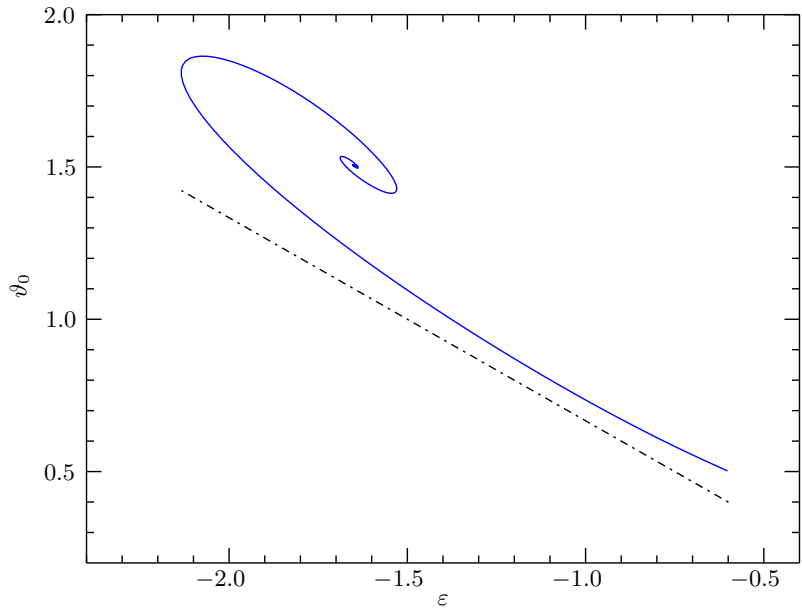}
\caption{Central dimensionless temperature $\vartheta_0$ of the King model as a function of the dimensionless energy $\varepsilon$ (blue solid line), compared to the caloric curve $\vartheta(\varepsilon)$ (black dash-dotted line) already reported in Fig.\ \protect\ref{figure:King_caloric}.}
\label{figure:spiral}
\end{figure}

The caloric curve of the King model is reported in Fig.\ \ref{figure:King_caloric}: the straight line is the virial relation (\ref{King_caloric_curve}), while the symbols are $(\varepsilon,\vartheta)$ values calculated using King distribution functions. 
The accumulation of points close to $(\varepsilon,\vartheta) = (\varepsilon^*,\vartheta^*) \simeq (-1.65,1.1)$ in Fig.\ \ref{figure:King_caloric} is a remarkable effect. Indeed, by calculating King distribution functions for increasing central densities $\varrho_0 = \varrho(r=0)$, the corresponding values of $\varepsilon$ decrease until $\varepsilon^{\text{King}}_{\text{min}}$ is reached. Continuing to increase $\varrho_0$ results in an oscillating pattern of $\varepsilon$ values centered around $\varepsilon = \varepsilon^*$, as in a ``collapsed spiral''. The value $\varepsilon = \varepsilon^*$ is asymptotically reached for $\varrho_0 \to \infty$. This behaviour closely reminds of the spiral behaviour of the caloric curve of the isothermal sphere: here the spiral collapses onto a straight line due to the constraint (\ref{King_caloric_curve}). To show that it is really a collapsed spiral, we plot not $\vartheta$ but a related quantity as a function of $\varepsilon$.  First, we define a ``central temperature'' $T_0$ by 
\beq
K_0 = \frac{3T_0}{2}
\eeq
where $K_0$ is the kinetic energy the system would have if $v^2$ were constant and equal to that in the center\footnote{This is a typical assumption made in the astrophysical context to estimate masses of gravitating systems from observed velocity dispersions, which are easier to measure in the central regions, see e.g.\ \cite{BinneyMerrifield:book}.}, $v^2_0 = v^2 (r = 0)$, 
\beq
K_0 = \frac{1}{2}M v_0^2\, .
\eeq
Then, in Fig.\ \ref{figure:spiral} we plot the dimensionless central temperature $\vartheta_0$ given by
\beq
\vartheta_0  =  \frac{r_t T_0}{GM^2} \label{King_central_theta}
\eeq 
as a function of $\varepsilon$. Having removed the constraint that the values must lie on a straight line, the spiral pattern opens up.

Although not expressed in terms of a caloric curve, the presence of a minimal value for the energy was already noticed in King's original paper \cite{King:aj1966_III} and interpreted as a stability limit for the model, suggesting that the model should be modified to accommodate more collapsed density profiles (the issue of the stability limits of King and other similar models was further elaborated in  \cite{Lynden-BellWood:mnras1968} and \cite{Katz:mnras1980}). From the above discussion it is apparent that the presence of a minimal energy $\varepsilon^{\text{King}}_{\text{min}}$ is nothing but the analogue in the King model context of the gravothermal catastrophe of the isothermal sphere. Hence, in view of what recalled in Sec.\ \ref{sec_modelcutoff}, we suggest that the addition of a small-scale cutoff to the interactions is likely to be a natural and physically motivated modification of the King model which stabilizes a low-energy phase where more collapsed density profiles are allowed.

The next Section is devoted to discussing the King model with a short-distance cutoff.  

\section{King model with a short-distance cutoff}
\label{Sec_King_cutoff}
Let us now define the King model with short-distance cutoff on the interactions. We assume that the distribution function is the King one (\ref{King_f}). We choose a Plummer short-distance regularization of the interactions, given by Eq.\ (\ref{cutoff}). The regularized gravitational potential $\varphi_\infty(\mathbf{r})$ generated by a mass density $\varrho$ can thus be written as 
\beq
\varphi_\infty(\mathbf{r}) = -G\int \frac{\varrho( \mathbf{r}' )}{\sqrt{\left|{\mathbf{r}'} - {\mathbf{r}}\right|^2 + a^2} }\, d\mathbf{r}'\, ;
\label{V_cutoff}
\eeq
as before, we denote by $\varphi_\infty$ the gravitational potential that vanishes at $r \to \infty$, to distinguish it from the potential $\varphi$ entering the expression (\ref{King_f}) of the distribution function, that vanishes at $r_t$. Since our density is spherically symmetric, also the potential must depend only on $r$. Indeed, for $\varrho( \mathbf{r} ) = \varrho( r )$ the angular integrations in Eq.\ (\ref{V_cutoff}) can be performed to yield
\beq
\varphi_\infty({r}) = -\frac{2\pi G}{r}\int_0^{r_t} r' \varrho( r' )\left[ \sqrt{(r'+r)^2 + a^2} - \sqrt{(r'-r)^2 + a^2}  \right]d{r}'\, ,
\label{V_cutoff_r}
\eeq
where we have taken into account that $\varrho( r ) = 0$ if $r > r_t$. The potential $\varphi( r )$ entering Eq.\ (\ref{King_f}) can be obtained from Eq.\ (\ref{V_cutoff_r}) by subtracting $\varphi_\infty(r_t)$ and reads as
\beq
\varphi({r}) = -{2\pi G}\int_0^{r_t} r' \varrho( r' )\left[ \frac{\sqrt{(r'+r)^2 + a^2} - \sqrt{(r'-r)^2 + a^2}}{r} - \frac{\sqrt{(r'+r_t)^2 + a^2} - \sqrt{(r'-r_t)^2 + a^2}}{r_t}   \right]d{r}'\, .
\label{phi_cutoff_r}
\eeq
Inspection of Eq.\ (\ref{phi_cutoff_r}) shows that $\varphi(r_t) = 0$. 

At variance with the standard King model, our potential does not obey a Poisson or Poisson-like equation. Hence to find the density profile $\varrho( r )$ explicitly we have to solve self-consistently the equation giving the density  as a function of the potential, that is Eq.\ (\ref{mfdensity}) with $f$ given by Eq.\ (\ref{King_f}), and the equation (\ref{phi_cutoff_r}) defining the potential as a function of the density. To this end it is convenient to define a dimensionless potential $W$ as
\beq
W( r ) = -2\gamma \varphi( r ) \, ,
\label{W}
\eeq
and an analogous $W_\infty$ defined as above with $\varphi_\infty$ in place of $\varphi$. In Eq.\ (\ref{W}) $\gamma$ is the inverse potential scale entering the definition (\ref{King_f}) of the distribution function. By definition, $W \geq 0$ if $r \leq r_t$ with $W(r_t) = 0$. We then integrate Eq.\ (\ref{mfdensity}) with $f$ given by Eq.\ (\ref{King_f}) to obtain
\beq
\varrho( r ) = \frac{4\pi C}{3\gamma^{3/2}} \Phi \left[ W( r )\right]\, ,
\label{rhoint}
\eeq
where
\beq
\Phi(W) = 2e^W \int_0^{\sqrt{W}} e^{-\eta^2} \eta^4 d\eta~.
\label{Phi}
\eeq
The normalization constant $C$ can be expressed as a function of the total mass $M$, 
\beq
C = \frac{3 \gamma^{3/2} M}{16 \pi^2 \int_0^{r_t} \Phi \left[ W( r ) \right] r^2 dr}~.
\eeq
Since $W(r_t) = 0$, Eq.\ (\ref{rhoint}) correctly implies $\varrho(r_t) = 0$. The two equations to be solved self-consistently are thus Eq.\ (\ref{rhoint}) and Eq.\ (\ref{W}), with $\varphi( r)$ given by Eq.\ (\ref{phi_cutoff_r}). It is convenient to rewrite them using dimensionless quantities. Let us put $r_t$ as unit of length, $M$ as unit of mass and $GM^2/r_t$ as unit of energy, in order to use the same units as in the previous Section. We then define a dimensionless radius $x$ as
\beq
x = \frac{r}{r_t}~,
\eeq
a dimensionless mass density $\psi$
\beq
\psi = \frac{\varrho\, r_t^3}{M}~,
\eeq
a dimensionless inverse potential scale $j^2$
\beq
j^2 = \frac{\gamma\, G M}{r_t}~,
\eeq
and a dimensionless cutoff length $\alpha$
\beq
\alpha = \frac{a}{r_t}~.
\eeq
Using these quantities the system of dimensionless equations to be numerically solved self-consistently can be written as
\bea
\psi(x) & = & \frac{\Phi[W(x)]}{4\pi \int_0^1 \Phi[W(y)] \, y^2 dy} \label{sc1}\\
W(x) & = & W_\infty(x) - W_\infty(1) \label{sc2}\\
W_\infty(x) & = & \frac{4\pi j^2}{x} \int_0^1 y\, \psi(y) \left[\sqrt{(y+x)^2+\alpha^2} - \sqrt{(y-x)^2+\alpha^2} \right] \, dy\label{sc3}
\eea
with $\Phi$ given by Eq.\ (\ref{Phi}). The above equations have to be solved iteratively, starting from an initial condition for the dimensionless density  $\psi_0(x)$. During the iterative procedure the constraints that the density and potential vanish at $x = 1$, i.e., $\psi(1) = W(1) = 0$, and
\beq
4 \pi \int_0^1 \psi(x) \, x^2 dx = 1~,
\label{m_constraint}
\eeq
i.e., that the total dimensionless mass is equal to one, must be satisfied. Due to the form of Eqs.\ (\ref{sc1}), (\ref{sc2}) and (\ref{sc3}) these constraints are automatically satisfied during the iterative procedure once an initial condition $\psi_0(x)$ compatible with the constraints is chosen. 

In practice, after fixing a value for $\alpha$, we choose a value $W_0$ and set $j^2$ such that $W(0) = W_0$; then we set $\psi$ to an initial form\footnote{A good choice for the initial density is $\psi_0(x) = \frac{3}{\pi}(1 - x)$.} $\psi_0(x)$  which satisfies the constraints. Then we compute $W$ from Eqs.\ (\ref{sc2}) and (\ref{sc3}) and we insert the function $W(x)$ thus obtained into Eq.\ (\ref{sc1}), finding a new dimensionless density $\psi(x)$ which is inserted again into Eq.\ (\ref{sc3}). We iterate the procedure until the integrated difference between densities at two subsequent steps is smaller than a prescribed threshold\footnote{We also check that the difference between physical observables like potential energy and kinetic energy evaluated at subsequent steps is smaller than a given threshold.}. This yields the desired dimensionless density profile corresponding to the chosen value of $W_0$.
Then we choose another value for the control parameter $W_0$ and repeat the procedure, starting from  the $\psi(x)$ found at the previous value of $W_0$: we also checked that the result does not change if we start from $\psi_0(x)$.

\subsection{Caloric curve}
\label{Sec_caloric_King_cutoff}

Given a density profile $\psi(x)$ and the corresponding potential profile $W(x)$ we can calculate the dimensionless kinetic and potential energy, from which we obtain the dimensionless temperature $\vartheta$ and energy $\varepsilon$.

The dimensional potential energy $U$ is 
\beq
U = 2\pi\int_0^{r_t}r^2\, \varphi_\infty( r ) \, \varrho ( r ) \, dr\, ,
\eeq
so that the dimensionless potential energy $u = Ur_t/(GM^2)$ is given by 
\beq
u = - \frac{\pi}{j^2}\int_0^1 W_\infty(x)\, \psi(x)\, x^2\, dx\, .
\label{u} 
\eeq
The dimensional kinetic energy $K$ can be written as
\beq
K = 8 \pi^2 \int_0^{r_t} dr \, r^2 \int_0^{v_e ( r )} dv\, v^4 \, f(r,v)\, ;
\eeq
with the change of variable $\eta = \gamma v^2$ and one partial integration we get
\beq
K = \frac{3M}{10\gamma \int_0^{r_t} \Phi[W( r )]\, r^2\, dr} \int_0^{r_t} dr \, r^2 \, e^{W ( r )} \int_0^{W ( r )} \eta^{5/2}\, e^{-\eta}\, d\eta\, ,
\eeq
hence the dimensionless kinetic energy $\kappa = K r_t/(GM^2)$ is
\beq
\kappa = \frac{3}{10 j^2 \int_0^{1} \Phi[W( x )]\, x^2\, dx} \int_0^{1} dx \, x^2 \, e^{W ( x )} \int_0^{W ( x )} \eta^{5/2}\, e^{-\eta}\, d\eta\, .
\label{kappa}
\eeq
The dimensionless temperature $\vartheta = T r_t/(GM^2)$ is then given by
\beq
\vartheta = \frac{2}{3} \kappa~,
\eeq
and the total dimensionless energy $\varepsilon$ is
\beq
\varepsilon = \kappa + u~.
\eeq
\begin{figure}
\psfrag{t}{$\vartheta$}
\psfrag{e}{$\varepsilon$}
\psfrag{-15}{\hspace{-0.2cm}$-15$}
\psfrag{-5}{\hspace{-0.2cm}$-5$}
\psfrag{-10}{\hspace{-0.2cm}$-10$}
\psfrag{0}{$0$}
\psfrag{0.0}{$0.0$}
\psfrag{0.5}{$0.5$}
\psfrag{1.0}{$1.0$}
\psfrag{1.5}{$1.5$}
\includegraphics[width=12cm,clip=true]{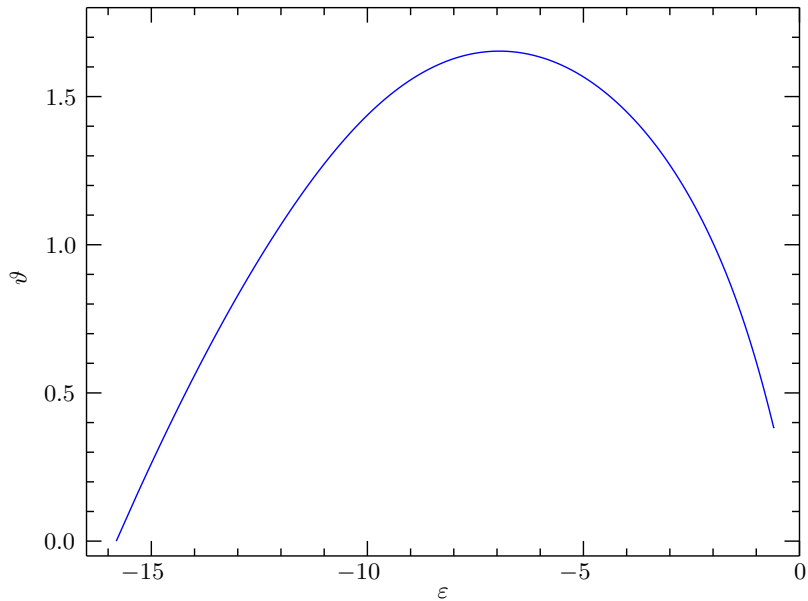}
\caption{Caloric curve of the King model with a short-distance cutoff. The dimensionless squared cutoff length is $\alpha^2 = 10^{-3}$.}
\label{figure:King_caloric_cutoff_large}
\end{figure}
\begin{figure}
\psfrag{t}{$\vartheta$}
\psfrag{e}{$\varepsilon$}
\psfrag{-150}{\hspace{-0.2cm}$-150$}
\psfrag{-100}{\hspace{-0.2cm}$-100$}
\psfrag{-50}{\hspace{-0.2cm}$-50$}
\psfrag{0}{$0$}
\psfrag{5}{$5$}
\psfrag{10}{$10$}
\psfrag{15}{$15$}
\includegraphics[width=12cm,clip=true]{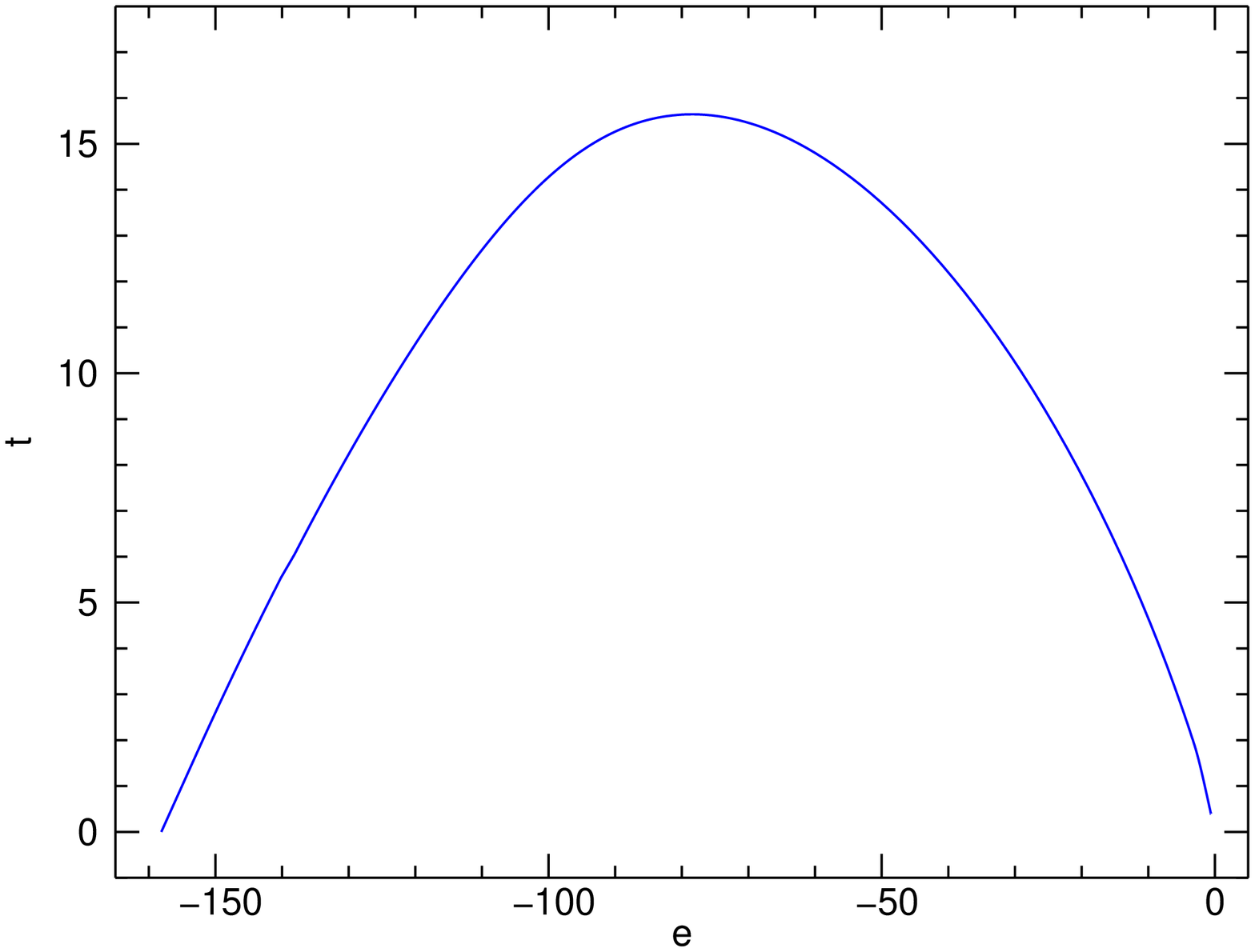}
\caption{As in Fig.\ \protect\ref{figure:King_caloric_cutoff_large}, with $\alpha^2 = 10^{-5}$.}
\label{figure:King_caloric_cutoff_medium}
\end{figure}
The caloric curve $\vartheta$ vs.\ $\varepsilon$ of the King model with short-distance cutoff obtained as explained above is reported in Fig.\ \ref{figure:King_caloric_cutoff_large} for a dimensionless squared cutoff length $\alpha^2 = 10^{-3}$ and in  Fig.\ \ref{figure:King_caloric_cutoff_medium} for a smaller value of the cutoff, $\alpha^2 = 10^{-5}$. Two facts are apparent: $(i)$ the small-scale cutoff stabilizes a low energy phase extending until energies well below the minimal energy $\varepsilon^{\text{King}}_{\text{min}} \simeq -2.13$, i.e., until the minimum value
\beq
\varepsilon^{\alpha}_{\text{min}} = u^{\alpha}_{\text{min}} = -\frac{1}{2\alpha}~ ,
\eeq
corresponding to the potential energy of a system where all the particles are in the same point, so that such a low-energy energy phase gets larger as the cutoff decreases; $(ii)$ the shape of the caloric curve is very similar to that of confined and regularized models, discussed in Sec.\ \ref{sec_modelcutoff} and sketched in Fig.\ \ref{figure:caloric}, but for the absence of the gas-like behaviour at high energies, which was allowed in those models by the presence of a confining box.

\begin{figure}
\psfrag{t}{$\vartheta$}
\psfrag{e}{$\varepsilon$}
\psfrag{-3.0}{\hspace{-0.2cm}$-3.0$}
\psfrag{-2.5}{\hspace{-0.2cm}$-2.5$}
\psfrag{-2.0}{\hspace{-0.2cm}$-2.0$}
\psfrag{-1.5}{\hspace{-0.2cm}$-1.5$}
\psfrag{-1.0}{\hspace{-0.2cm}$-1.0$}
\psfrag{-0.5}{\hspace{-0.2cm}$-0.5$}
\psfrag{0.5}{$0.5$}
\psfrag{1.0}{$1.0$}
\psfrag{1.5}{$1.5$}
\psfrag{2.0}{$2.0$}
\includegraphics[width=12cm,clip=true]{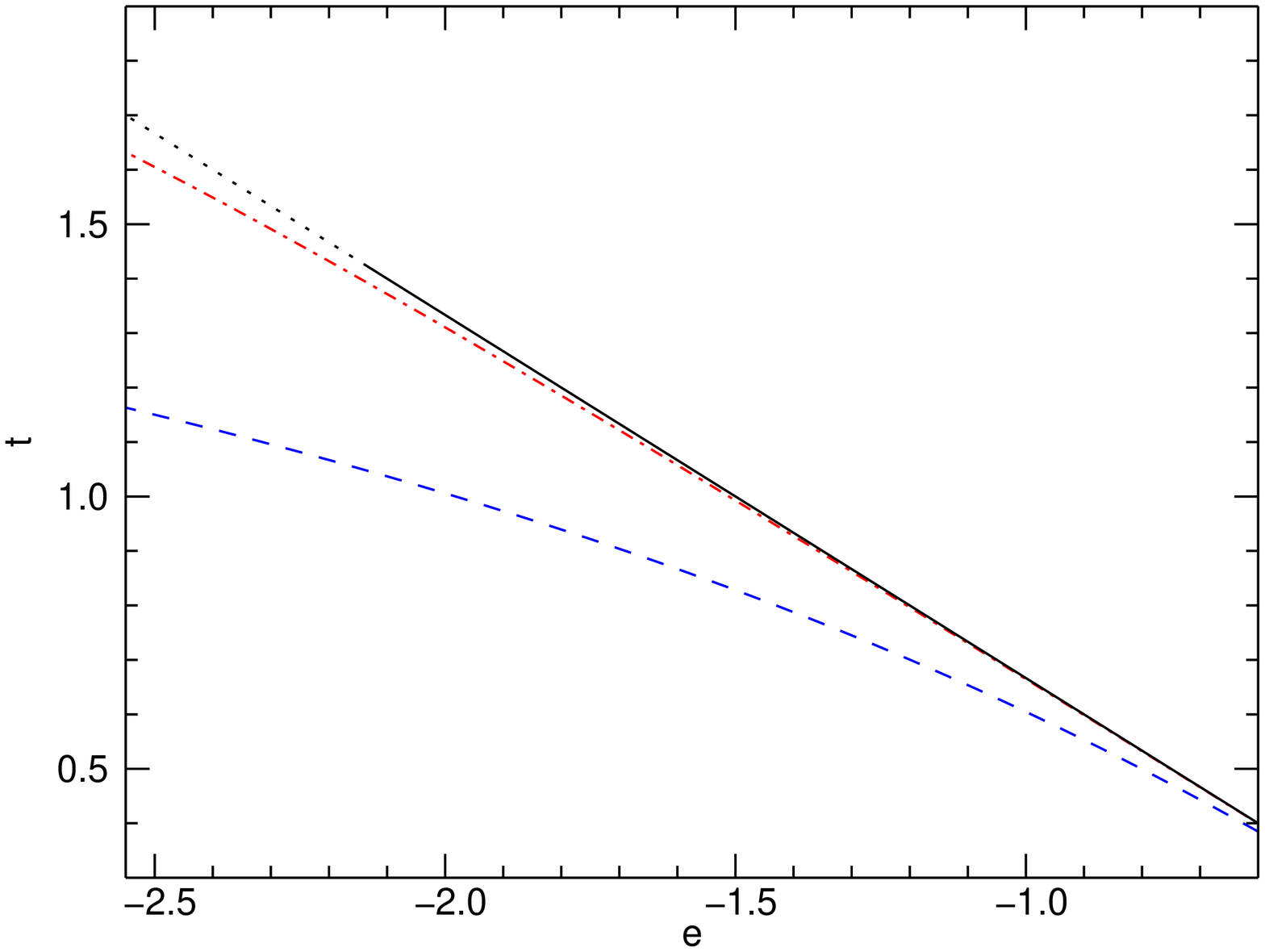}
\caption{Caloric curves of the King model with a short-distance cutoff (blue dashed line:  $\alpha^2 = 10^{-3}$, red dot-dashed line:  $\alpha^2 = 10^{-5}$) compared with the caloric curve of the King model without cutoff (black solid line). The dotted line is the continuation of the virial law $\vartheta = -\frac{2}{3}\varepsilon$ for energies below $\varepsilon^{\text{King}}_{\text{min}} \simeq -2.13$.}
\label{figure:King_caloric_cutoff_comparison}
\end{figure}

In Fig.\ \ref{figure:King_caloric_cutoff_comparison} a close-up of the caloric curves with cutoff (for the two cutoff lengths $\alpha^2 = 10^{-3}$ and  $\alpha^2 = 10^{-5}$ of Figs.\ \ref{figure:King_caloric_cutoff_large} and \ref{figure:King_caloric_cutoff_medium}, respectively) and without cutoff is shown in energy interval containing the minimal energy of the King model without cutoff, $\varepsilon^{\text{King}}_{\text{min}} \simeq -2.13$. We observe that already for $\alpha^2 = 10^{-5}$ the caloric curve with cutoff is practically the same as that without cutoff, where the latter exists, but extends well below the latter as already shown before. 

These results give a first and positive answer to the question we put at the beginning of the present work, i.e., whether the behaviour of regularized and confined models of self-gravitating systems might be relevant to real systems where neither confinement nor proper  thermal equilibrium states exist. Indeed, by studying a short-distance regularized version of an observationally probed model, the King model, we have shown that the caloric curve of such a model is qualitatively the same as that of regularized and confined model, the only qualitative difference being the absence, in the King model with cutoff, of a gas-like phase at high energies. This difference was expected from the outset since the gas-like phase is allowed only due to the confinement. Moreover, we have shown that already for moderate values of the cutoff ($\alpha^2 = 10^{-5}$, see Fig.\ \ref{figure:King_caloric_cutoff_comparison}) the caloric curve of the King model with cutoff is virtually indistinguishable from that of the original King model without cutoff, for values of energies where both curves exist.

In the next two subsections we study the shape of the density profiles obtained with the King model with cutoff and we address the problem of understanding what happens when the cutoff gets smaller than the values considered up to now.

\subsection{Density profiles}
\label{Sec_profiles_King_cutoff}

We can recognize three different regions in the caloric curve of the King model with cutoff. First, a high-energy region, allowed also for the model without cutoff, for values of $\epsilon \gtrsim \varepsilon^{\text{King}}_{\text{min}}$. Second, an intermediate-energy region, which is the continuation of the caloric curve without cutoff for $\epsilon \lesssim \varepsilon^{\text{King}}_{\text{min}}$ and where the specific heat remains negative. Third, a low-energy region where the specific heat becomes positive and is the cutoff-dominated region. 

\begin{figure}
\psfrag{x}{$x$}
\psfrag{p}{$\psi(x)$}
\psfrag{1.e-06}{\hspace{0.15cm}$10^{-6}$}
\psfrag{1.e-04}{\hspace{0.15cm}$10^{-4}$}
\psfrag{1.e-02}{\hspace{0.15cm}$10^{-2}$}
\psfrag{1.e+00}{\hspace{0.4cm}$10^{0}$}
\psfrag{1.e+02}{\hspace{0.4cm}$10^{2}$}
\psfrag{1.e+04}{\hspace{0.4cm}$10^{4}$}
\psfrag{0.01}{$\hspace{0.05cm}0.01$}
\psfrag{0.10}{$\hspace{0.05cm}0.1$}
\psfrag{1.00}{$\hspace{0.2cm}1$}
\includegraphics[width=12cm,clip=true]{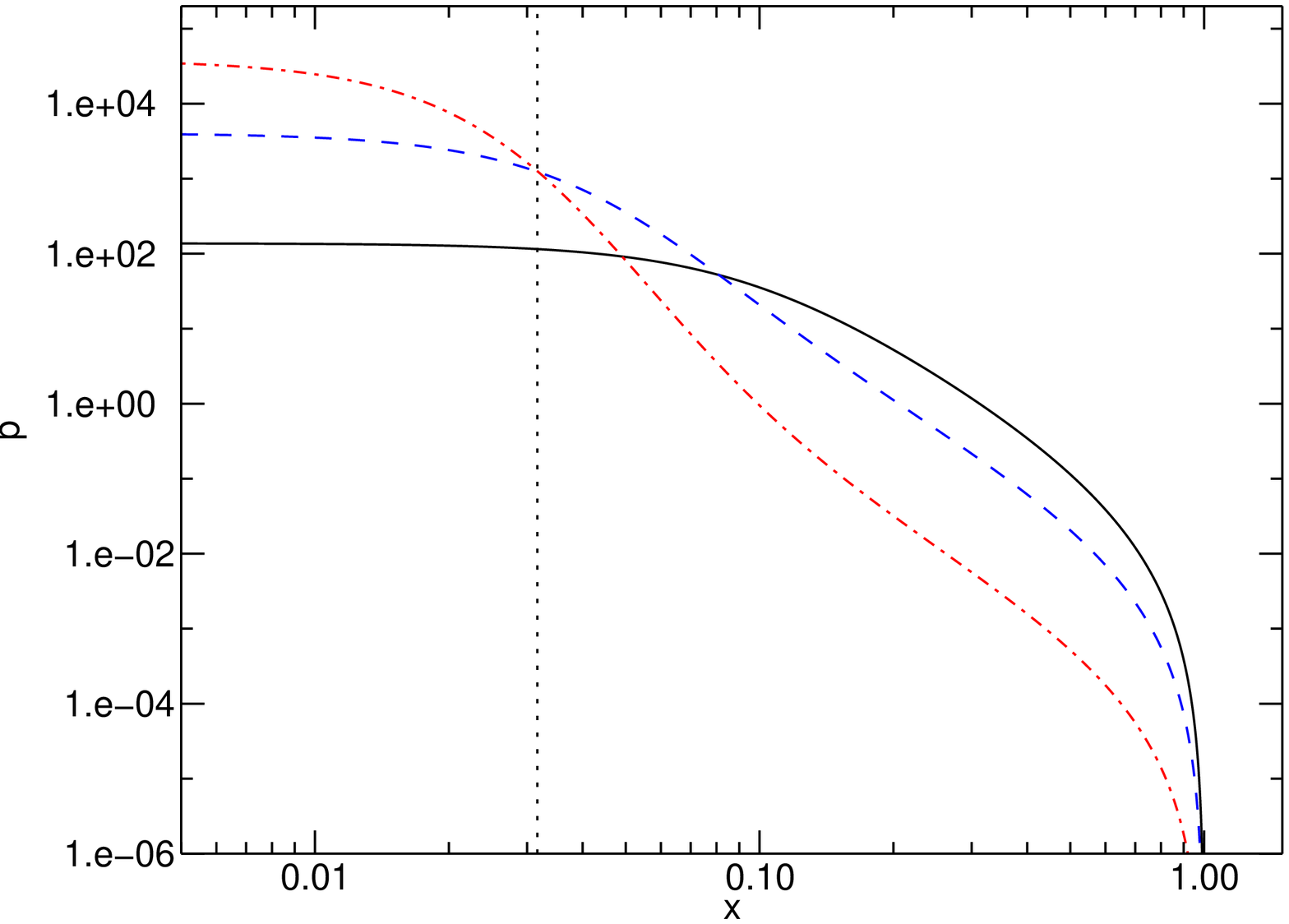}
\caption{Dimensionless density profile $\psi(x)$ as a function of the dimensionless radius $x$ for the King model with short-distance cutoff, with $\alpha^2 = 10^{-3}$. Black solid curve: $\varepsilon = -1.27$, belonging to the energy range in common with the King model without cutoff; blue dashed curve: $\varepsilon = -3.91$, belonging to the intermediate energy region with negative specific heat; red dot-dashed curve: $\varepsilon = -9.02$, belonging to the lowest energy region with positive specific heat. The dotted vertical line marks the cutoff length scale $\alpha$.}
\label{figure:psi10m3}
\end{figure}

\begin{figure}
\psfrag{x}{$x$}
\psfrag{p}{$\psi(x)$}
\psfrag{1.e-06}{\hspace{0.15cm}$10^{-6}$}
\psfrag{1.e-04}{\hspace{0.15cm}$10^{-4}$}
\psfrag{1.e-02}{\hspace{0.15cm}$10^{-2}$}
\psfrag{1.e+00}{\hspace{0.4cm}$10^{0}$}
\psfrag{1.e+02}{\hspace{0.4cm}$10^{2}$}
\psfrag{1.e+04}{\hspace{0.4cm}$10^{4}$}
\psfrag{1.e+06}{\hspace{0.4cm}$10^{6}$}
\psfrag{1.e+08}{\hspace{0.4cm}$10^{8}$}
\psfrag{0.001}{$\hspace{0.05cm}0.001$}
\psfrag{0.010}{$\hspace{0.15cm}0.01$}
\psfrag{0.100}{$\hspace{0.15cm}0.1$}
\psfrag{1.000}{$\hspace{0.3cm}1$}
\includegraphics[width=12cm,clip=true]{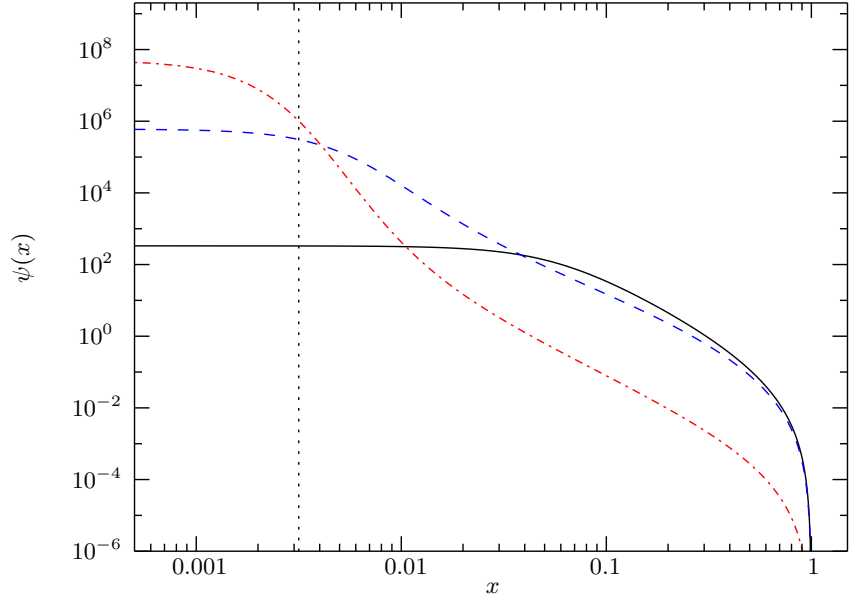}
\caption{As in Fig.\ \protect\ref{figure:psi10m3}, with $\alpha^2 = 10^{-5}$ and energies:  $\varepsilon = -1.37$ (black solid curve); $\varepsilon = -7.19$ (blue dashed curve); $\varepsilon = -96.4$ (red dot-dashed curve).}
\label{figure:psi10m5}
\end{figure}

It is interesting to have a look at the density profiles $\psi(x)$ corresponding to the three different regions. Examples of density profiles are reported in Figs.\ \ref{figure:psi10m3} and \ref{figure:psi10m5} for $\alpha^2 = 10^{-3}$ and $\alpha^2 = 10^{-5}$, respectively. Note that the density profiles are plotted using log-log scales. The main feature emerging from these plots is that there is a qualitative change in the shape of the density profile passing from the first to the second and third regions of energy. In the high-energy region, where for $\alpha^2 = 10^{-5}$ the caloric curves with cutoff and without cutoff are nearly indistinguishable, the density profile is, as expected, very close to the density profile of the King model without cutoff\footnote{We checked that King density profiles without cutoff and with a cutoff $\alpha^2 = 10^{-5}$ agree also quantitatively.}, i.e., it starts with a flattened core and then decays without inflections. When we lower the energy an inflection point in the profile appears and a central core develops, whose density grows as the energy is lowered. For the smaller of the two cutoffs, the change between the two kinds of density profiles (without and with the central core) is much more abrupt. 

The interest of this phenomenon is twofold. First, it suggests that the physics in the energy region that was not allowed in the model without cutoff is not a simple continuation of the previous one and that at smaller cutoff the change between the two kinds of density profiles might correspond to a phase-transition-like phenomenon; this possibility will be addressed in the next Section. Second, it has also a phenomenological interest. We have mentioned that the King model provides good fits to nearly 80\% of the Milky Way globular clusters. Most of the remaining 20\% are referred to as ``(post-)core-collapsed'' clusters \cite{DjorgovskiKing:apjl1986}. It is believed that, after having undergone a gravothermal-like collapse in the central region which has been stopped by dynamical effects like the formation of binary stars, these clusters have settled in a near-equilibrium state which cannot be described by the King model because their density profile is too much concentrated \cite{MeylanHeggie:aareview1997}. Many of these clusters have  surface brightness profiles with a central core, qualitatively similar\footnote{To make a comparison one should project our density profiles onto a plane to obtain a surface density, since it is the surface brightness profile that is observed; anyhow, the shape of the profiles would not change that much in the projection, so that the qualitative observation is still valid.} to those we obtain with our model with cutoff. A detailed comparison with observations is outside the scope of the present work, and we are not claiming our model with cutoff is able to provide a good fit for the density profiles of core-collapsed clusters; however, the qualitative similarity between observed and theoretical profiles suggests that the introduction of a short-distance cutoff in the interactions might be a possible way to construct simple models able to describe also these clusters\footnote{We would also like to note that recent observations made with the Hubble Space Telescope \cite{NoyolaGebhardt:aj2006,NoyolaGebhardt:aj2007} have shown that also in some globular clusters whose ground-based observations were well fitted by the King model without cutoff an excess central density exists.}. 


\subsection{Phase transition for small values of the cutoff?}
\label{Sec_King_small_cutoff}

The analysis reported in Sec.\ \ref{Sec_profiles_King_cutoff} has shown that in the low energy regime, i.e., when $\varepsilon \lesssim \varepsilon^{\text{King}}_{\text{min}}$, the density profiles of the King model with short-distance cutoff are qualitatively different from those at higher energies, the latter being essentially equal to those of the King model without cutoff. This suggests that for small $\alpha$'s there might be a sharp crossover or even a singularity in the caloric curve for $\varepsilon \simeq \varepsilon^{\text{King}}_{\text{min}}$. This would be analogous to the phase transition between different states with negative specific heat that is observed e.g.\ in the isothermal sphere with short-distance cutoff \cite{Chavanis:ijmpb2006}, for small values of the cutoff.

There is another reason to expect that something special happens to the caloric curve close to $\varepsilon \simeq \varepsilon^{\text{King}}_{\text{min}}$,  for small values of $\alpha$. The caloric curve of the King model with cutoff must become equal to that of the King model without cutoff when $\alpha \to 0$. We have already seen that, for energies in the domain of the caloric curve without cutoff, the caloric curve with cutoff becomes practically equal to that without cutoff already for $\alpha^2 = 10^{-5}$, so that we expect this to be true for any $\alpha^2 \lesssim 10^{-5}$. However, for finite $\alpha$ the domain of the caloric curve with cutoff appears much larger than that without cutoff, and getting larger as $\alpha$ gets smaller; but when $\alpha \to 0$ also the domains of the caloric curves must become the same, so that all the low energy region $\varepsilon < \varepsilon^{\text{King}}_{\text{min}}$ ought to disappear. There could be, in principle, (at least) two different scenarios leading to that. In the first scenario the limit $\alpha \to 0$ is a singular limit: the whole low-energy region exists for any $\alpha > 0$ and abruptly disappears when $\alpha \to 0$. In this case one could also expect that for sufficiently small $\alpha$ a phenomenon analogous to an equilibrium phase transition exists, where the caloric curve develops a singularity separating the low-energy phase from the high-energy phase whose domain contains (and when $\alpha \to 0$ coincides with) that of the King model without cutoff. In the second scenario, for sufficiently small $\alpha$ a forbidden region in energy $\varepsilon_{\text{low}}(\alpha) < \varepsilon < \varepsilon_{\text{high}}(\alpha)$ opens up, where no solution of the self-consistent equations (\ref{sc1}), (\ref{sc2}) and (\ref{sc3}) exists. The extrema of the forbidden region in energy would be such that $\lim_{\alpha\to 0}\varepsilon_{\text{low}}(\alpha) = \varepsilon^\alpha_{\text{min}}$ and $\lim_{\alpha\to 0}\varepsilon_{\text{high}}(\alpha) = \varepsilon^{\text{King}}_{\text{min}}$, so that although the minimum energy decreases with $\alpha$, also the effective width of the low energy phase decreases and eventually shrinks to a point as $\alpha \to 0$. 

\begin{figure}
\psfrag{t}{$\vartheta$}
\psfrag{e}{$\varepsilon$}
\psfrag{-8}{\hspace{-0.2cm}$-8$}
\psfrag{-6}{\hspace{-0.2cm}$-6$}
\psfrag{-4}{\hspace{-0.2cm}$-4$}
\psfrag{-2}{\hspace{-0.2cm}$-2$}
\psfrag{1}{$1$}
\psfrag{2}{$2$}
\psfrag{3}{$3$}
\psfrag{4}{$4$}
\includegraphics[width=12cm,clip=true]{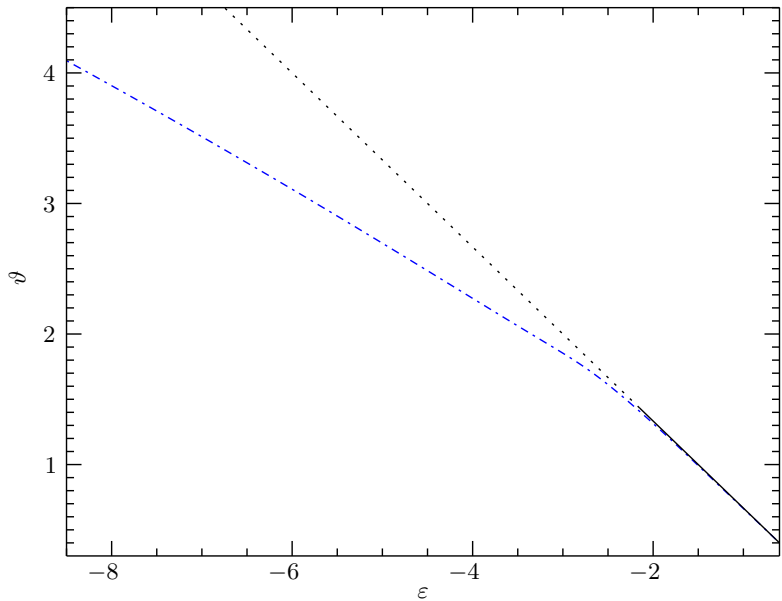}
\caption{Comparison between the caloric curve of the King model with a short-distance cutoff with $\alpha^2 = 7.5\times 10^{-6}$ (blue dot-dashed line) and the caloric curve of the King model without cutoff (black solid line). The dotted line is the continuation of the virial law $\vartheta = -\frac{2}{3}\varepsilon$ for energies below $\varepsilon^{\text{King}}_{\text{min}} \simeq -2.13$. A change in the slope of the caloric curve with cutoff is apparent for energies between 2 and 3.}
\label{figure:transition}
\end{figure}
\begin{figure}
\psfrag{W0}{$W_0$}
\psfrag{e}{$\varepsilon$}
\psfrag{6}{$6$}
\psfrag{7}{$7$}
\psfrag{8}{$8$}
\psfrag{9}{$9$}
\psfrag{10}{$10$}
\psfrag{11}{$11$}
\psfrag{12}{$12$}
\psfrag{0}{$0$}
\psfrag{-10}{\hspace{-0.2cm}$-10$}
\psfrag{-20}{\hspace{-0.2cm}$-20$}
\psfrag{-30}{\hspace{-0.2cm}$-30$}
\psfrag{-40}{\hspace{-0.2cm}$-40$}
\includegraphics[width=12cm,clip=true]{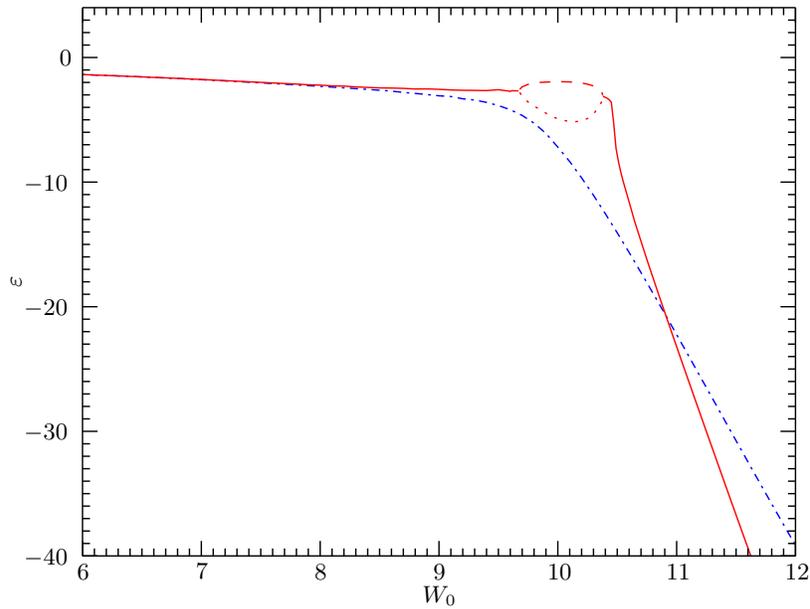}
\caption{Dimensionless energy $\varepsilon$ as a function of the control parameter $W_0$ (dimensionless central potential) for the King model with a short-distance cutoff. Blue dot-dashed curve: $\alpha^2 = 10^{-5}$; red solid curve: $\alpha^2 = 5 \times 10^{-6}$, values of $\varepsilon$ corresponding to converged (non-oscillating) solutions; red dashed and dotted curves: $\alpha^2 = 5 \times 10^{-6}$, values of $\varepsilon$ corresponding to upper and lower bounds of oscillating solutions.}
\label{figure:w0eps}
\end{figure}
\begin{figure}
\psfrag{t}{$\vartheta$}
\psfrag{e}{$\varepsilon$}
\psfrag{-6}{\hspace{-0.2cm}$-6$}
\psfrag{-5}{\hspace{-0.2cm}$-5$}
\psfrag{-4}{\hspace{-0.2cm}$-4$}
\psfrag{-3}{\hspace{-0.2cm}$-3$}
\psfrag{-2}{\hspace{-0.2cm}$-2$}
\psfrag{-1}{\hspace{-0.2cm}$-1$}
\psfrag{1.0}{$1.0$}
\psfrag{1.5}{$1.5$}
\psfrag{2.0}{$2.0$}
\psfrag{2.5}{$2.5$}
\psfrag{3.0}{$3.0$}
\psfrag{3.5}{$3.5$}
\includegraphics[width=12cm,clip=true]{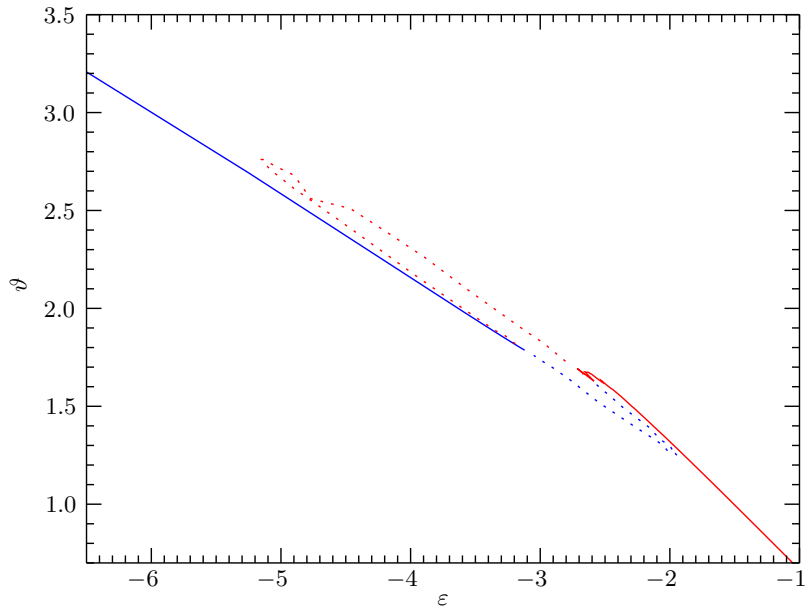}
\caption{Caloric curve of the King model with a short-distance cutoff, in an anergy range similar to that of Fig.\ \ref{figure:transition}. The dimensionless squared cutoff length is $\alpha^2 = 5 \times 10^{-6}$. The red and blue solid lines are converged values, while the red and blue dotted lines refer to the lower and upper bounds of oscillating solutions.}
\label{figure:King_caloric_cutoff_small}
\end{figure}

The results we have obtained so far do not allow to discriminate between the two scenarios yet, although the phase-transition-like scenario seems favoured. The first result suggesting a phase transition might occur is shown in Fig.\ \ref{figure:transition}, where we plot the caloric curve for $\alpha^2 = 7.5 \times 10^{-6}$,  in a slightly wider energy window than in Fig.\ \ref{figure:King_caloric_cutoff_comparison}, and a change of slope in $\vartheta(\varepsilon)$ at energies between 2 and 3 is apparent, which reminds of a caloric curve near a continuous phase transition. This suggests that for slightly smaller $\alpha$ the caloric curve may develop a kink and, for even smaller $\alpha$, a discontinuity like in a first-order phase transition. However, upon slightly decreasing the cutoff length to $\alpha^2 = 5 \times 10^{-6}$, we were not able to find a reliable convergence to a self-consistent solution for values of the control parameter such that $9.67 < W_0 < 10.38$. When trying to solve the self-consistent equations for these values of $W_0$, the solution did not converge yet also after a number of iterations orders of magnitude larger than the number of iterations leading to convergence in the rest of the curve. In these cases the kinetic and potential energies oscillate between two different values. In Fig.\ \ref{figure:w0eps} the values of $\varepsilon$ are plotted as a function of the control parameter $W_0$, for two choices of the cutoff, $\alpha^2 = 10^{-5}$ and $\alpha^2 = 5 \times 10^{-6}$. In the latter case, in the region where no reliable convergence was found, upper and lower bounds of the oscillations are reported.
This behaviour is compatible with both scenarios, because oscillating solutions may indicate either true absence of self-consistent solutions or a hysteresis-like phenomenon, due to the existence of metastable states, which in turn might suggest a first-order-transition-like behaviour. Indeed, even if no self-consistent solutions exists for $W_0 \in[W_0^{\text{low}}, W_0^{\text{high}}]$, this does not automatically mean that a forbidden region in $\varepsilon$ exists; this would require also that $\varepsilon(W_0^{\text{low}}) \not = \varepsilon(W_0^{\text{high}})$, while the shape of the curve reported in Fig.\ \ref{figure:w0eps} suggests that $\varepsilon(W_0^{\text{low}}) = \varepsilon(W_0^{\text{high}})$ is not ruled out. The caloric curve $\vartheta(\varepsilon)$ for $\alpha^2 = 5 \times 10^{-6}$ is plotted in Fig.\ \ref{figure:King_caloric_cutoff_small}; solid lines refer to converged solutions while dotted lines are upper and lower bounds of oscillating solutions. Although a small window in energy where no converged solutions exist is present ($-3.120 < \varepsilon < -2.715$), bounds of oscillating solutions lie on continuations of the converged solutions and resemble metastable states typically observed close to discontinuous phase transitions. Indeed, finding the converged solution for energies where also oscillating solutions exists required a fine sampling of $W_0$ values as well as a very large number of iterations: with the typical number of iterations leading to convergence in other regions of the curve, one would see oscillations. Therefore, at least for $\alpha^2 = 5\times 10^{-6}$, the presence of the small energy region without any converged solutions might well be a purely numerical problem due to the very slow convergence of our numerical method in this region, which could perhaps be solved employing a much larger number of iterations \footnote{We performed the calculations using both \textsc{Mathematica} and \textsc{Matlab} packages, using adaptive sampling of the functions with up to $10^4$ points and $10^4$ iterations; some of the calculations have also been checked with an independently written C code.}. Hence the phase-transition-like scenario seems favoured. 
Going to even smaller values of $\alpha$ yields a very similar behaviour, but for that the $W_0$ range where solutions oscillate gets larger. 

\begin{figure}
\psfrag{x}{$x$}
\psfrag{p}{$\psi(x)$}
\psfrag{1.e-06}{\hspace{0.15cm}$10^{-6}$}
\psfrag{1.e-04}{\hspace{0.15cm}$10^{-4}$}
\psfrag{1.e-02}{\hspace{0.15cm}$10^{-2}$}
\psfrag{1.e+00}{\hspace{0.4cm}$10^{0}$}
\psfrag{1.e+02}{\hspace{0.4cm}$10^{2}$}
\psfrag{1.e+04}{\hspace{0.4cm}$10^{4}$}
\psfrag{1.e+06}{\hspace{0.4cm}$10^{6}$}
\psfrag{1.e+08}{\hspace{0.4cm}$10^{8}$}
\psfrag{0.001}{$\hspace{0.05cm}0.001$}
\psfrag{0.010}{$\hspace{0.15cm}0.01$}
\psfrag{0.100}{$\hspace{0.15cm}0.1$}
\psfrag{1.000}{$\hspace{0.3cm}1$}
\includegraphics[width=12cm,clip=true]{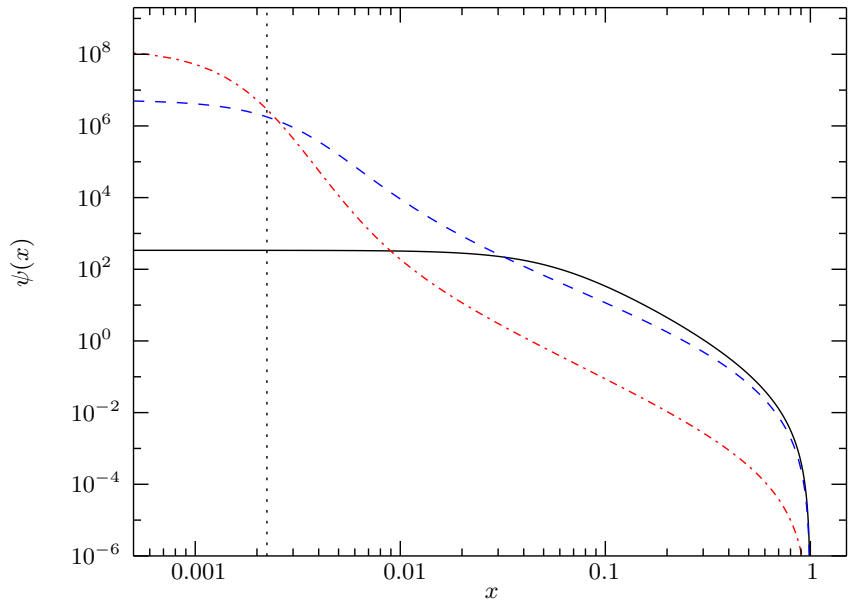}
\caption{As in Fig.\ \protect\ref{figure:psi10m3}, with $\alpha^2 = 5\times 10^{-6}$ and energies:  $\varepsilon = -1.37$ (black solid curve); $\varepsilon = -17.65$ (blue dashed curve); $\varepsilon = -136$ (red dot-dashed curve).}
\label{figure:psi5x10m6}
\end{figure}

The overall shape of the caloric curve for $\alpha^2 = 5 \times 10^{-6}$ is very similar to that shown in Fig.\ \ref{figure:King_caloric_cutoff_medium} so that we do not show it. 
Some density profiles obtained with $\alpha^2 = 5 \times 10^{-6}$ are plotted in Fig.\ \ref{figure:psi5x10m6}. We observe that the qualitative features of the profiles are the same as for larger cutoffs. Hence going to smaller values of $\alpha$ does not qualitatively change the shape of the density profiles.

\section{Concluding remarks}
\label{Sec_final}

The main motivation of the present work was to try to understand whether the behaviour of regularized and confined models of self-gravitating systems previously studied in equilibrium statistical mechanics could be considered relevant to real self-gravitating systems like globular star clusters, where there is no confinement so that the stationary velocity distribution is not the thermal one. To this end, we have studied a modified King model of a globular cluster, with a short-distance cutoff on the interactions. We have shown that such a model does behave like the regularized and confined ones, especially for what concerns the caloric curve, but for the absence of a high-energy gas-like phase which can only be induced by confinement. The caloric curve and density profiles of our model become equal to those of the observationally probed King model without cutoff for cutoff lengths such that $\alpha^2 \lesssim 10^{-5}$, for energy values larger than the minimum energy allowed for the King model without cutoff. 

Our results thus provide a first, positive answer to the question that motivated our work.

There are still many open points. First of all, the issue of the presence and of the nature of a transitional phenomenon for small cutoffs, discussed in Sec.\ \ref{Sec_King_small_cutoff}, should be clarified. To this end it might probably be useful to consider different regularization schemes allowing for numerically easier treatments (possibly in a differential form). Moreover, although the model studied in the present work does not have a proper thermal equilibrium, one might construct an effective entropy in close analogy to Eq.\ (\ref{mfentropy}) and try to use it to study the stability of the solutions.

Another open problem concerns the physical interpretation of the small-scale cutoff. In the present work, as in previous works recalled in Sec.\ \ref{sec_modelcutoff}, the short-distance cutoff has been taken as an external parameter, whose presence is required by physical considerations and whose numerical values should be sufficiently small to yield results not too far from those of the King model without cutoff at large energies and are otherwise arbitrary. However, physical considerations should also provide reasonable bounds for the cutoff scale or even estimates of its numerical value. Clearly the cutoff scale must be bounded below by the size of the individual stars; an order of magnitude for an upper bound could be given by the average interstellar distance. Translated into our dimensionless units this means $10^{-9} \lesssim \alpha \lesssim 5 \times 10^{-2}$, so that all the cutoffs we have considered above fit into this range. Theoretical considerations suggest that the cutoff scale is more likely to be close to the upper bound than to the lower bound: since we describe the system with a single-particle distribution function, the cutoff scale should be of the same order of magnitude as the interparticle distance in order to be consistent with neglecting correlation effects (see e.g.\ \cite{NelsonTremaine:mnras1999}). On the other hand, the cutoff might be considered as an effective implementation of dynamical effects that stabilize an otherwise unstable highly-concentrated distribution of mass, like the formation of binary stars: reasonable values of the cutoff may thus be suggested by analyzing such effects. 

As a final remark we recall that the King model with cutoff allows for energies much smaller than those of the King model without cutoff, corresponding to density profiles with a central core. From a phenomenological point of view, this suggests that also core-collapsed globular clusters not well described by the King model might be accommodated in a description with cutoff. A detailed comparison between the model predictions and observed density profiles is left for future work: in case the density profiles with cutoff turn out to yield reasonable descriptions of core-collapsed clusters, these clusters might be interpreted as the manifestation of a phase transition between different structures described by the same model, instead of as something requiring a completely different description. 

\acknowledgments
We warmly thank A.\ Marconi for enlightening discussions and for his interest in our work, and A.\ Patelli for checking some of our numerical results with an independently written C code. CN acknowledges support from the EGIDE scholarship funded by Minist\`ere des Affaires \'etrang\`eres (France).

\bibliography{/Users/casetti/Work/Scripta/papers/bib/mybiblio,/Users/casetti/Work/Scripta/papers/bib/astro,/Users/casetti/Work/Scripta/papers/bib/statmech}

\end{document}